\documentclass[journal]{IEEEtran}
\usepackage{amsfonts}
\usepackage{cite,graphicx,amsmath,amsthm}
\allowdisplaybreaks[4]
\usepackage{subfigure}
\usepackage{fancyhdr}
\usepackage{dsfont}
\usepackage{array,color}
\usepackage{bm}
\usepackage{booktabs}
\usepackage{multirow}
\usepackage{algorithm}
\usepackage{algpseudocode}

\usepackage{bbm}
\usepackage{graphicx}

\newtheorem{theorem}{Theorem}

\newtheorem{lemma}{Lemma}

\newtheorem{proposition}{Proposition}

\newtheorem{corollary}{Corollary}

\newtheorem{property}{Property}

\newtheorem{remark}{Remark}

\newtheorem{claim}{Claim}

\begin{document}
\title{{Backscatter Data Collection with Unmanned Ground Vehicle: Mobility Management\\and Power Allocation}}

\author{Shuai Wang, Minghua Xia, and Yik-Chung Wu      
\thanks{
This work was supported by the National Natural Science Foundation of China (NSFC) under Grant No. 61671488, and by the Major Science and Technology Special Project for ``New Generation Communication and Network'', Guangdong Province, China.

S. Wang and Y.-C. Wu are with the Department of Electrical and Electronic Engineering, The University of Hong Kong, Hong Kong (e-mail: \{swang, ycwu\}@eee.hku.hk).

M. Xia is with the School of Electronics and Information Technology, Sun Yat-sen University, Guangzhou, 510006, China (e-mail: xiamingh@mail.sysu.edu.cn).
}
}

\maketitle

\begin{abstract}
Collecting data from massive Internet of Things (IoT) devices is a challenging task, since communication circuits are power-demanding while energy supply at IoT devices is limited.
To overcome this challenge, backscatter communication emerges as a promising solution as it eliminates radio frequency components in IoT devices. Unfortunately, the transmission range of backscatter communication is short.
To facilitate backscatter communication, this work proposes to integrate unmanned ground vehicle (UGV) with backscatter data collection.
With such a scheme, the UGV could improve the communication quality by approaching various IoT devices.
However, moving also costs energy consumption and a fundamental question is: what is the right balance between spending energy on moving versus on communication?
To answer this question, this paper studies energy minimization under a joint graph mobility and backscatter communication model.
With the joint model, the mobility management and power allocation problem unfortunately involves nonlinear coupling between discrete variables brought by mobility and continuous variables brought by communication.
Despite the optimization challenges, an algorithm that theoretically achieves the minimum energy consumption is derived, and it leads to automatic trade-off between spending energy on moving versus on communication in the UGV backscatter system.
Simulation results show that if the noise power is small (e.g., $\leq-100~\mathrm{dBm}$), the UGV should collect the data with small movements. However, if the noise power is increased to a larger value (e.g., $-60~\mathrm{dBm}$), the UGV should spend more motion energy to get closer to IoT users.
\end{abstract}


\IEEEpeerreviewmaketitle
\section{Introduction}

\IEEEPARstart{W}ith a wide range of commercial and industrial applications, Internet of Things (IoT) market is continuously growing \cite{iot1}, and the number of inter-connected IoT devices is expected to exceed 20 billion by 2020.
However, these massive IoT devices (e.g., sensors and tags) are usually limited in size and energy supply \cite{wpt}, making data collection challenging in IoT systems.
To this end, backscatter communication is a promising solution, because it eliminates radio frequency (RF) components in IoT devices \cite{back1,back2,back3,back4,back5}.
Unfortunately, due to the round-trip path-loss, the transmission range of backscatter communication is limited \cite{back6,back7,back8}.
This can be seen from a recent prototype in \cite{back1}, where the wirelessly powered backscatter communication only supports a range of $1$ meter at the data-rate of $1$ $\mathrm{kbps}$.

To combat the short communication range, this paper investigates a viable solution that the backscatter RF transmitter and tag reader are mounted on an unmanned ground vehicle (UGV).
With such a scheme, the UGV could vary its location for wireless data collection, thus having the flexibility of being close to different IoT devices at different times \cite{ugv2}.
However, since moving the UGV would consume motion energy, an improperly chosen path might lead to excessive movement, thus offseting the benefit brought by movement \cite{ugv3,ugv4,ugv5,ugv6}.
Therefore, the key is to balance the trade-off between spending energy on moving versus on communication, which unfortunately cannot be handled by traditional vehicle routing algorithms \cite{laporte1,ugv7,ugv8}, since they do not take the communication power and quality-of-service (QoS) into account.

In view of the apparent research gap, this paper proposes an algorithm that leads to automatic trade-off in spending energy on moving versus on communication.
In particular, the proposed algorithm is obtained by integrating the graph mobility model and the backscatter communication model.
With the proposed model, the joint mobility management and power allocation problem is formulated as a QoS constrained energy minimization problem.
Nonetheless, such a problem turns out to be a mixed integer nonlinear programming problem (MINLP), which is nontrivial to solve due to the nonlinear coupling between discrete variables brought by moving and continuous variables brought by communication.
This is in contrast to unmanned aerial vehicle (UAV) based systems in which only continuous variables are involved \cite{uav1,uav2,uav3,uav4,uav5}.
To this end, the optimality condition of the MINLP is first established, which helps in reducing the problem dimension.
Then, an efficient algorithm, which is guaranteed to obtain the global optimal solution, is proposed.
By adopting the proposed algorithm, minimum energy consumption is achieved at the UGV, and simulation results are presented to further demonstrate the performance of the proposed algorithm.

The rest of this paper is organized as follows. In Section II, the system model, which includes the mobility model and the backscatter communication model, is described.
Then, the joint mobility management and power allocation problem is formulated in Section III.
The algorithm for computing the optimal solution is derived in Section IV, and an efficient initialization is proposed in Section V.
Finally, numerical results are presented in Section VI, and conclusions are drawn in Section VII.

\emph{Notation}.
Italic letters, simple bold letters, and capital bold letters represent scalars, vectors, and matrices, respectively.
Curlicue letters represent sets and $|\cdot|$ is the cardinality of a set.
We use $(a_1,a_2,\cdots)$ to represent a sequence and $[a_1,a_2,\cdots]^{T}$ to represent a column vector, with $(\cdot)^{T}$ being the transpose operator.
The operators $\textrm{Tr}(\cdot)$ and $(\cdot)^{-1}$ take the trace and the inverse of a matrix, respectively.
Finally, $\mathbb{E}(\cdot)$ represents the expectation of a random variable.

\section{System Model}

\subsection{Mobility Model}
\begin{figure}[!t]
\centering
\includegraphics[width=50mm]{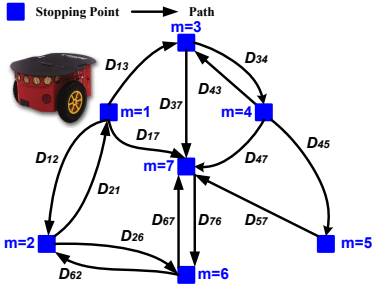}
\caption{UGV mobility model with $M=7$.}
\label{fig_sim}
\end{figure}

\setcounter{secnumdepth}{4}We consider a wireless data collection system, which consists of $K$ IoT users and one UGV equipped with a RF transmitter and a tag reader.
The environment in which the UGV operates in is described by a directed graph $(\mathcal{V},\mathcal{E})$ as shown in Fig. 1, where $\mathcal{V}=\{1,\cdots,M\}$ is the set of $M$ vertices representing the possible stopping points, and $\mathcal{E}$ is the set of directed edges representing the allowed movement paths \cite{graph}.
To quantify the path length, a matrix $\mathbf{D}=[D_{1,1},\cdots,D_{1,M};\cdots;D_{M,1},\cdots,D_{M,M}]\in\mathbb{R}^{M\times M}_+$ is defined, with the element $D_{m,j}$ representing the distance from vertex $m$ to vertex $j$ ($D_{m,m}=0$ for any $m$).
If there is no allowed path from vertex $m$ to vertex $j$, we set $D_{m,j}=+\infty$ \cite{graph}.
Notice that $\mathbf{D}$ is a constant matrix since the locations of all vertices are pre-determined.

To model the movement of the UGV, we define a visiting path $\mathcal{Q}=(y_1,y_2,\cdots,y_Q)$ where $y_j \in \mathcal{V}$ for $j=1,\cdots,Q$ and $(y_j,y_{j+1})\in \mathcal{E}$ for $j=1,\cdots,Q-1$, with $Q-1$ being the number of steps to be taken.
Without loss of generality, we assume the following two conditions hold:
\begin{itemize}
\item[(i)] $y_1=y_Q$. This is generally true as a typical UGV management scenario is to have the UGV standing by at the starting point (e.g., for charging and maintenance services) after the data collection task  \cite{laporte1}.
    For notational simplicity, it is assumed that vertex $y_1=y_Q=1$ is the start and end point of the path to be designed.

\item[(ii)] There are no repeating vertices among $(y_1,\cdots,y_{Q-1})$. This is true because if a vertex $m$ is visited twice, we can always introduce an auxiliary vertex with $D_{M+1,j}=D_{m,j}$ and $D_{j,M+1}=D_{j,m}$ for all $j\in\mathcal{V}$  \cite{laporte1,laporte2}.
    Thus this scenario can be represented by an extended graph with one more vertex and an extended $\mathbf{D}$ with dimension $(M+1)\times(M+1)$.

\end{itemize}

Correspondingly, we define the selection variable $\mathbf{v}=[v_1,\cdots,v_M]^{T}\in\{0,1\}^M$, where $v_m=1$ if the vertex $m$ appears in the path $\mathcal{Q}$ and $v_m=0$ otherwise.
Furthermore, we define a matrix $\mathbf{W}=[W_{1,1},\cdots,W_{1,M};\cdots;W_{M,1},\cdots,W_{M,M}]\in\{0,1\}^{M\times M}$, with $W_{y_j,y_{j+1}}=1$ for all $j=1,\cdots,Q-1$ and zero otherwise.

With the moving time from the vertex $m$ to the vertex $j$ being $D_{m,j}/a$ where $a$ is the velocity,
the total moving time along path $\mathcal{Q}$ is
\begin{align}
\frac{1}{a}\sum_{m=1}^M\sum_{j=1}^MW_{m,j}D_{m,j}=\frac{\mathrm{Tr}(\mathbf{D}^{T}\mathbf{W})}{a}. \label{SW}
\end{align}
Furthermore, since the total motion energy $E$ of the UGV is proportional to the total motion time \cite{ugv2,ugv3,ugv4}, the motion energy can be expressed in the form of
\begin{align}
E=\left(\frac{\alpha_1}{a}+\alpha_2\right)\mathrm{Tr}(\mathbf{D}^{T}\mathbf{W}), \label{EW}
\end{align}
where $\alpha_1$ and $\alpha_2$ are parameters of the model (e.g., for a Pioneer 3DX robot in Fig. 1, $\alpha_1=0.29$ and $\alpha_2=7.4$ \cite[Sec. IV-C]{ugv2}).

\subsection{Backscatter Data Collection Model}

Based on the mobility model, the UGV moves along the selected path $\mathcal{Q}$ to collect data from users as shown in Fig. 2.
In particular, from the starting point $y_1$, the UGV stops for a duration $u_{y_1}$ and then it moves along edge $(y_1,y_2)$ to its outward neighbor $y_2$, and stops for a duration $u_{y_2}$. The UGV keeps on moving and stopping along the path until it reaches the destination $y_Q$.

When the UGV stops at the vertex $m$ (with $v_m=1$), it will wait for a time duration $u_m$ for data collection.
Out of this $u_m$, a duration of $t_{k,m}$ will be assigned to collect data from user $k$ via full-duplex backscatter communication\footnote{When user $k$ adapts the variable impedance for modulating the backscattered waveform with information bits, other users keep silent to avoid collision \cite{back4}.} \cite{FD1,FD2}. More specifically, if $t_{k,m}=0$, the IoT user $k$ will not be served in duration $u_m$.
On the other hand, if $t_{k,m}\neq0$, the RF source at the UGV transmits a symbol $x_{k,m}\in\mathbb{C}$ with $\mathbb{E}[|x_{k,m}|^2]=p_{k,m}$, where $p_{k,m}$ is the transmit power of the RF source.
Then the received signal-to-noise ratio (SNR) at the UGV tag reader is
$\eta |g_{k,m}|^2|h_{k,m}|^2 p_{k,m}/N_0$, where $h_{k,m}\in \mathbb{C}$ is the downlink channel from the UGV to user $k$, $g_{k,m}\in \mathbb{C}$ is the uplink channel from user $k$ to the UGV, and $N_0$ is the power of complex Gaussian noise (including the self-interference due to full-duplex backscatter \cite{FD1,FD2}).
Furthermore, $\eta$ is the tag scattering efficiency determined by the load impedance $Z_L$ and the antenna impedance $Z_{A}$ \cite{back10}.
For example, in the on-off keying backscatter shown in Fig. 2, the IoT device switches between two load impedances $Z_1$ and $Z_2$ with $Z_1\neq Z_A$ and $Z_2=Z_A$.
This means that the IoT device transmits $(Z_L-Z_{A})/(Z_L+Z_{A})=(Z_1-Z_{A})/(Z_1+Z_{A})$ when switching to $Z_L=Z_1$ and transmits $(Z_L-Z_{A})/(Z_L+Z_{A})=0$ when switching to $Z_L=Z_2$.
Therefore, $\eta=|(Z_1-Z_{A})/(Z_1+Z_{A})|^2$ in the on-off keying.

Based on the backscatter model, the transmission rate during $t_{k,m}$ is given by
\begin{align}\label{rate}
R_{k,m}=\mathrm{log}_2\left(1+v_m\cdot\frac{\beta\eta|g_{k,m}|^2|h_{k,m}|^2 p_{k,m}}{N_0}\right),
\end{align}
where $\beta$ is the performance loss due to imperfect modulation and coding schemes in backscatter communication \cite{back11}.
For example, in bistatic backscatter communication with frequency shift keying, $\beta=0.5$ \cite{back11}.
On the other hand, in ambient backscatter communication with on-off keying, $\beta$ is obtained by fitting $\mathrm{log}_2\left(1+\beta x\right)$ to $1-\mathbb{Q}\left(\sqrt{x}\right)$ \cite{ook}, where
$\mathbb{Q}\left(x\right)=1/\sqrt{2\pi}\int_x^\infty \mathrm{exp}\left(-u^2/2 \right)\mathrm{d}u$ refers to the Q-function.

\begin{figure}[!t]
\centering
\includegraphics[width=70mm]{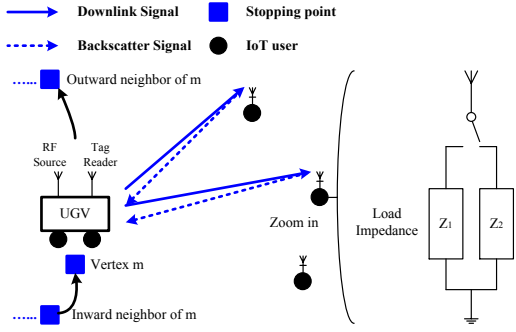}
\caption{Backscatter data collection operation with UGV.}
\label{fig_sim}
\end{figure}

\emph{Remark 1:}
The channels $\{g_{k,m},h_{k,m}\}$ can be pre-determined as follows.
If the environment is static, all the information about object positions, geometry and dielectric properties in the environment is available.
In such a case, ray tracing methods \cite{E1} could be used to estimate $\{g_{k,m},h_{k,m}\}$.
On the other hand, if the channel is varying but with a fixed distribution, we could allow the UGV to collect a small number of measurements at the stopping points before a set of new missions (e.g., five missions) \cite{E3}.
Then based on the probabilistic framework in \cite{E3}, the UGV can predict the channels at all the stopping points.

\emph{Remark 2:}
In practice, the backscatter efficiency would vary with the incident power.
However, if the change of backscatter efficiency is not significant, it is possible to adopt a constant $\eta$ to facilitate the analysis \cite{back6,back7,back11,back13}.
On the other hand, if $\eta$ is not a constant, according to \cite{back12}, it is possible to compute the range of $\eta$.
Then, the worst case approach, which replaces $\eta$ in \eqref{rate} with its lower bound $\eta_{\mathrm{lb}}$, can be adopted to guarantee the data collection targets for all users.
Notice that this worst case approach can only achieve suboptimal performance, and modeling $\eta$ as a nonlinear function of the incident power is an important future work.

\section{Joint Mobility Management and Power Allocation}

In wireless data collection systems, the task is to collect certain amount of data from different IoT devices by planning the path (involving variables $\mathbf{v}$ and $\mathbf{W}$) and designing the stopping time $\{t_{k,m}\}$ and transmit power $\{p_{k,m}\}$.
In particular, the data collection QoS requirement of the $k^{\mathrm{th}}$ IoT device can be described by
\begin{align}
&\sum_{m=1}^M
t_{k,m}\cdot\mathrm{log}_2\left(1+v_m\cdot\frac{\beta\eta|g_{k,m}|^2|h_{k,m}|^2 p_{k,m}}{N_0}\right)\geq\gamma_k,
\end{align}
where $\gamma_k>0$ (in $\mathrm{bit/Hz}$) is the amount of data to be collected from user $k$.

Notice that the variables $\mathbf{v}$ and $\mathbf{W}$ are dependent since $v_m=0$ implies $W_{m,j}=W_{j,m}=0$ for any $j\in \mathcal{V}$.
On the other hand, the UGV would visit the vertex with $v_m=1$, making $\sum_{j=1}^M W_{m,j} = \sum_{j=1}^M W_{j,m}=1$.
Combining the above two cases, we have
\begin{align}
&\sum_{j=1}^MW_{m,j}=v_m,~\sum_{j=1}^MW_{j,m}=v_m,~~\forall m=1,\cdots,M.
\end{align}
Furthermore, since the path must be connected, the following subtour elimination constraints are required to eliminate disjointed sub-tours \cite{laporte2}:
\begin{align}
&\lambda_m-\lambda_j+\left(\sum_{l=1}^Mv_l-1\right)W_{m,j}+\left(\sum_{l=1}^Mv_l-3\right)W_{j,m}
\nonumber\\
&\leq \sum_{l=1}^Mv_l-2
+J\left(2-v_m-v_j\right)
,~~\forall m,j\geq 2,~m\neq j,
\nonumber\\
&v_m\leq\lambda_m\leq\left(\sum_{l=1}^Mv_l-1\right)v_m,~~\forall m\geq 2,
\end{align}
where $\{\lambda_{m}\}$ are slack variables to guarantee a connected path, and $\sum_{l=1}^Mv_l$ is the number of vertices involved in the path.
The constant $J=10^{6}$ is large enough such that the first line of constraint is always satisfied when $v_m=0$ or $v_j=0$.
In this way, the vertices not to be visited would not participate in subtour elimination constraints.

Having the data collection and graph mobility constraints satisfied, it is then crucial to reduce the total energy consumption at the UGV.
As the energy consumption includes motion energy
$\left(\alpha_1/a+\alpha_2\right)\mathrm{Tr}(\mathbf{D}^{T}\mathbf{W})$
and communication energy $\sum_{m=1}^M\sum_{k=1}^Kt_{k,m}p_{k,m}$, the joint mobility management and power allocation problem of the data collection system is formulated as\footnote{The considered system adopts semi-passive backscatter communication, where the local circuits at IoT devices are powered by their own batteries \cite{back3}. On the other hand, if the users also request data from the UGV, simultaneous wireless information and power transfer from the UGV to IoT users can be adopted \cite{Claessens1,Claessens2,Claessens3,Claessens4,Claessens5}.}:
\begin{subequations}
\begin{align}
\mathrm{P}1:
&\mathop{\mathrm{min}}_{\substack{\mathbf{v},\mathbf{W},\{\lambda_m\}\\ \{t_{k,m},p_{k,m}\}}}
~~\mu\left(\frac{\alpha_1}{a}+\alpha_2\right)\mathrm{Tr}(\mathbf{D}^{T}\mathbf{W})
\nonumber\\
&~~~~~~~~~~~~~~
+(2-\mu)\sum_{m=1}^M\sum_{k=1}^Kt_{k,m}p_{k,m}
\nonumber
\\\mathrm{s.t.}~&\sum_{m=1}^M
t_{k,m}\cdot\mathrm{log}_2\left(1+v_m\cdot\frac{\beta\eta|g_{k,m}|^2|h_{k,m}|^2 p_{k,m}}{N_0}\right)
\nonumber\\
&
\geq\gamma_k,~~\forall k, \label{P1a}
\\
&\frac{1}{a}\mathrm{Tr}(\mathbf{D}^{T}\mathbf{W})+\sum_{m=1}^M\sum_{k=1}^Kt_{k,m}\leq T, \label{P1b}
\\
&
\sum_{j=1}^MW_{m,j}=v_m,~\sum_{j=1}^MW_{j,m}=v_m,~~\forall m, \label{P1c}
\\
&\lambda_m-\lambda_j+\left(\sum_{l=1}^Mv_l-1\right)W_{m,j}+\left(\sum_{l=1}^Mv_l-3\right)W_{j,m}
\nonumber\\
&
\leq \sum_{l=1}^Mv_l-2+J\left(2-v_m-v_j\right),
\nonumber\\
&
\forall m,j\geq 2,~m\neq j, \label{subtour1}
\\
&v_m\leq\lambda_m\leq\left(\sum_{l=1}^Mv_l-1\right)v_m,~~\forall m\geq 2, \label{subtour2}
\\
&W_{m,j}\in\{0,1\},~~\forall m,j,~~W_{m,m}=0,~~\forall m, \label{edge}
\\
&v_{1}=1,~v_{m}\in\{0,1\},~~\forall m\geq 2, \label{vertex}
\\
&(1-v_m)\cdot t_{k,m}=0,~~\forall k,m, \label{notvisit}
\\
&t_{k,m}\geq0,~p_{k,m}\geq0,~~\forall k,m, \label{resource}
\end{align}
\end{subequations}
where \eqref{P1b} is for constraining the operation (including moving and data collection) to be completed within $T$ seconds, and \eqref{notvisit} is for constraining the stopping time to be zero if the vertex is not visited.
Notice that $0<\mu\leq 1$ is a weighting factor to control the relative importance between motion energy and communication energy.
Nominally, if we are only interested in minimizing the total energy, we can set $\mu=1$.
On the other hand, if we want to restrict the interference to other co-existing wireless systems, we might set $\mu<1$.

It can be seen from the constraint \eqref{P1a} of $\mathrm{P}1$ that the UGV can choose the stopping vertices, which in turn affect the channel gains to and from the IoT users.
By choosing the stopping vertices with better channel gains to IoT users, the transmit powers $\{p_{k,m}\}$ might be reduced.
However, this might also lead to additional motion energy, which in turn costs more energy consumption at the UGV.
Therefore, there exists a trade-off between moving and communication, and solving $\mathrm{P}1$ can concisely balance this energy trade-off.

Unfortunately, problem $\mathrm{P}1$ is nontrivial to solve due to the following reasons.
Firstly, it is NP-hard, since it involves the integer constraints \eqref{edge}$-$\eqref{vertex} \cite{dimitri}.
Secondly, the data-rate and the energy cost at each vertex are dependent on the transmit power $\{p_{k,m}\}$ and transmission time $\{t_{k,m}\}$, which are unknown (see Table I).
This is in contrast to traditional integer programming problems \cite{dimitri}, where the reward of visiting each vertex is a constant.

\begin{table*}[ht]
\caption{Summary of Symbol Notations} 
\centering 
\begin{tabular}{|l|l|l|}
\hline
\textbf{Variable} & \textbf{Description} \\
\hline
$v_{m}\in\{0,1\}$  & $v_{m}=1$ represents the vertex $m$ being involved in the path; $v_{m}=0$ otherwise.\\
$W_{m,j}\in\{0,1\}$ & $W_{m,j}=1$ represents the edge $(m,j)$ being involved in the path; $W_{m,j}=0$ otherwise.  \\
$t_{k,m}\in \mathbb{R}_+$ & Time (in $\mathrm{s}$) allocated to user $k$ when UGV is at the $m^{\mathrm{th}}$ stopping point.  \\
$p_{k,m}\in \mathbb{R}_+$ & Transmit power (in $\mathrm{Watt}$) to user $k$ when UGV is at the $m^{\mathrm{th}}$ stopping point.  \\
\hline
\textbf{Parameter} & \textbf{Description} \\
\hline
$M$ & Number of stopping points. \\
$K$ & Number of users. \\
$\mathcal{V},\mathcal{E}$ & $\mathcal{V}$ ($\mathcal{E}$) is the set of all vertices (edges). \\
$D_{m,j}$ & Distance (in $\mathrm{m}$) from the $m^{\mathrm{th}}$ vertex to the $j^{\mathrm{th}}$ vertex. \\
$\mu$ & Weighting factor of motion energy. \\
$\alpha_1,\alpha_2$ & Parameters of the UGV motion energy model. \\
$a$ & Constant velocity (in $\mathrm{m/s}$) of the UGV. \\
$g_{k,m},h_{k,m}$ & $g_{k,m}$ ($h_{k,m}$) is the downlink (uplink) channel between the $m^{\mathrm{th}}$ vertex and the $k^{\mathrm{th}}$ user. \\
$T$ & Completion time (in $\mathrm{s}$) of the data collection and moving along the path. \\
$\beta$ & Performance loss due to imperfect modulation and coding schemes. \\
$\eta$ & Tag scattering efficiency. \\
$N_0$ & Receiver noise power (in $\mathrm{Watt}$). \\
$\gamma_k$ & The communicaiton QoS target (in $\mathrm{bit/Hz}$) at IoT user $k$. \\
\hline
\end{tabular}
\end{table*}

\section{Optimal Solution to $\mathrm{P}1$}

Despite the optimization challenges, this section proposes an algorithm that theoretically obtains the optimal solution to $\mathrm{P}1$.
The idea of this algorithm is to eliminate the variables $\mathbf{W},\{\lambda_m\},\{t_{k,m},p_{k,m}\}$ so as to transform $\mathrm{P}1$ into an equivalent problem only related to $\mathbf{v}$.
By doing so, we can capitalize on the branch and bound (B$\&$B) method and obtain the optimal solution by pruning out impossible candidates.
In the following, the optimality condition of $\mathrm{P}1$ will be first discussed, which helps in reducing the dimension of $\mathrm{P}1$.

\subsection{Optimality Condition}

To address the challenges for solving $\mathrm{P}1$, we first establish the optimality condition of $\mathrm{P}1$.
In particular, by defining
\begin{align}\label{Akm}
&A_{k,m}=\frac{\beta\eta|g_{k,m}|^2|h_{k,m}|^2}{N_0},
\end{align}
the following proposition (proved in Appendix A) can be established.
\begin{proposition}
The optimal $\{\mathbf{v}^*,t^*_{k,m},p_{k,m}^*\}$ to $\mathrm{P}1$ satisfies:

\noindent (i) If $m=\mathop{\mathrm{argmax}}_{l\in\mathcal{V}}~v_l^*A_{k,l}$, then $t^*_{k,m}\neq0$; otherwise $t^*_{k,m}=0$.

\noindent (ii) If $t_{k,m}^*\neq 0$, then $p^*_{k,m}\neq0$.

\end{proposition}

\textbf{Proposition 1} indicates that the UGV only needs to allocate time to user $k$ at a single vertex, which is given by $m=\mathrm{argmax}_{l\in\mathcal{V}}~v^*_lA_{k,l}$.
For other vertices, the allocated time to user $k$ should be zero.
Based on part (i) of \textbf{Proposition 1}, we can set the transmit time
\begin{align}
&t_{k,m}=\left\{
\begin{aligned}
&s_k
,&\mathrm{if}~m=\mathop{\mathrm{argmax}}_{l\in\mathcal{V}}~v_lA_{k,l}&
\\
&0,&\mathrm{if}~m\neq\mathop{\mathrm{argmax}}_{l\in\mathcal{V}}~v_lA_{k,l}&
\end{aligned}
\right.,
\label{tkm}
\end{align}
where $s_k>0$, without changing the optimal solution to $\mathrm{P}1$.
Correspondingly, the transmit power $\{p_{k,m}\}$ can be set to\footnote{If $m\neq\mathop{\mathrm{argmax}}_{l\in\mathcal{V}}~v_lA_{k,l}$, we have $t_{k,m}=0$. Thus $t_{k,m}p_{k,m}=0$ in the objective of $\mathrm{P}1$ and
$t_{k,m}\mathrm{log}_2\left(1+v_mA_{k,m}p_{k,m}\right)=0$ in \eqref{P1a}, meaning that $p_{k,m}$ would not participate in problem $\mathrm{P}1$.
As a result, we can set $p_{k,m}=0$.}
\begin{align}
&p_{k,m}=\left\{
\begin{aligned}
&q_k
,&\mathrm{if}~m=\mathop{\mathrm{argmax}}_{l\in\mathcal{V}}~v_lA_{k,l}&
\\
&0,&\mathrm{if}~m\neq\mathop{\mathrm{argmax}}_{l\in\mathcal{V}}~v_lA_{k,l}&
\end{aligned}
\right.,
\label{pkm}
\end{align}
where $q_k$ is the transmit power corresponding to $s_k$.
Since $s_k>0$, by part (ii) of \textbf{Proposition 1}, we also have $q_k>0$.
Putting \eqref{tkm} and \eqref{pkm} into $\mathrm{P}1$, problem $\mathrm{P}1$ is transformed into
\begin{subequations}
\begin{align}
\mathrm{P}2:
\mathop{\mathrm{min}}_{\substack{\mathbf{v},\mathbf{W},\{\lambda_m\}\\ \{s_{k}>0,q_{k}>0\}}}
~~&\mu\left(\frac{\alpha_1}{a}+\alpha_2\right)\mathrm{Tr}(\mathbf{D}^{T}\mathbf{W})
\nonumber\\
&
+(2-\mu)\sum_{k=1}^Ks_kq_{k}
\nonumber\\
~~~\mathrm{s.t.}~~~~~~&s_{k}\cdot\mathrm{log}_2\left[1+\left(\mathop{\mathrm{max}}_{l\in\mathcal{V}}~v_lA_{k,l}\right)q_k\right]
\nonumber\\
&
\geq\gamma_k,~~\forall k, \label{P2a}
\\
&\frac{\mathrm{Tr}(\mathbf{D}^{T}\mathbf{W})}{a}+\sum_{k=1}^Ks_{k}\leq T, \label{P2b}
\\
&\eqref{P1c}-\eqref{vertex}. \label{P2c}
\end{align}
\end{subequations}
Notice that  the constraint \eqref{notvisit} is dropped since \eqref{notvisit} is always satisfied when $\{t_{k,m}\}$ takes the form of \eqref{tkm}.

The problem $\mathrm{P}2$ is still nontrivial to solve due to the nonlinear coupling between $\mathbf{v}$ and $\{\mathbf{W},s_{k},q_{k}\}$ as observed from the constraints \eqref{P2a} and \eqref{P2c}.
To resolve such coupling, a straightforward idea is to use alternating minimization for optimizing $\mathbf{v}$, $\mathbf{W}$ and $\{s_{k},q_{k}\}$ iteratively.
However, due to the discrete nature of $\mathbf{v}$ and $\mathbf{W}$, such a method could fail to converge.
To this end, this paper proposes to simplify the problem based on elimination of variables.
In particular, we will first derive the optimal solution of $\mathbf{W}$ and $\{s_{k},q_{k}\}$ to $\mathrm{P}2$ with fixed $\mathbf{v}$.
By representing $\mathbf{W}$ and $\{s_{k},q_{k}\}$ as functions of $\mathbf{v}$, problem $\mathrm{P}2$ is simplified to an equivalent problem only involving $\mathbf{v}$.
Then we will step further to find the optimal solution of vertex selection variable $\mathbf{v}$.

\subsection{Optimal Solution of $\mathbf{W}$ and $\{s_{k},q_{k}\}$ with Fixed $\mathbf{v}$}

When $\mathbf{v}=\widetilde{\mathbf{v}}$, where $\widetilde{\mathbf{v}}$ is any feasible solution to $\mathrm{P}2$, the constraint \eqref{vertex} can be dropped since it only involves $\mathbf{v}$.
On the other hand, the term $\mathop{\mathrm{max}}_{l\in\mathcal{V}}~\widetilde{v}_l A_{k,l}$ in constraint \eqref{P2a} becomes a constant, and we denote it as $B_k(\widetilde{\mathbf{v}}):=\mathop{\mathrm{max}}_{l}~\widetilde{v}_l A_{k,l}$.
Furthermore, it can be seen from the objective function of $\mathrm{P}2$ that $q_k$ is a variable to be minimized and $s_{k}\mathrm{log}_2\left[1+B_k(\widetilde{\mathbf{v}})q_k\right]$ is a strictly increasing function of $q_k$.
As a result, the optimal solution of $q_k$ must activate the constraint \eqref{P2a} of $\mathrm{P}2$, which leads to
\begin{align}
&q_k=\frac{1}{B_k(\widetilde{\mathbf{v}})}\left(2^{\gamma_k/s_k}-1\right). \label{qk}
\end{align}
Putting \eqref{qk} into $\mathrm{P}2$, it is proved in Appendix B that the optimal $\mathbf{W}^*$ and $\{s_k^*\}$ to $\mathrm{P}2$ must activate the constraint \eqref{P2b}, i.e., $\mathrm{Tr}(\mathbf{D}^{T}\mathbf{W}^*)/a+\sum_{k=1}^Ks_{k}^*=T$.
Using this result, the quantity $\mathrm{Tr}(\mathbf{D}^{T}\mathbf{W})$ in the objective function of $\mathrm{P}2$ can be replaced by $a(T-\sum_{k=1}^Ks_{k})$ without changing the problem, and the objective function would be independent of $\mathbf{W}$.
Then problem $\mathrm{P}2$ is equivalently transformed into the following two-stage optimization problem (detailed procedure given in Appendix B):
\begin{align}
\mathrm{P}3:
\mathop{\mathrm{min}}_{\substack{\{s_{k}>0\}}}
~&\mu\left(\alpha_1+\alpha_2a\right)\left(T-\sum_{k=1}^Ks_{k}\right)
\nonumber\\
&
+(2-\mu)\sum_{k=1}^K\frac{\gamma_k}{B_k(\widetilde{\mathbf{v}})}
\Theta\left(\frac{s_k}{\gamma_k}\right)
\nonumber\\
~~\mathrm{s.t.}~~~&\sum_{k=1}^Ks_{k}
=\mathop{\mathrm{max}}_{\substack{\mathbf{W},\{\lambda_m\}}} \Big\{T-\frac{\mathrm{Tr}(\mathbf{D}^{T}\mathbf{W})}{a}:
\nonumber\\
&~~~~~~~~~~~~
\eqref{P1c}-\eqref{edge}\Big\},\nonumber
\end{align}
where $\Theta(x):=x(2^{1/x}-1)$.

To solve $\mathrm{P}3$, we first need to compute the right hand side of the constraint, which leads to the following problem:
\begin{align}
\mathop{\mathrm{max}}_{\substack{\mathbf{W},\{\lambda_{m}\}}}
~T-\frac{\mathrm{Tr}(\mathbf{D}^{T}\mathbf{W})}{a},~~~~~
\mathrm{s.t.}~\eqref{P1c}-\eqref{edge}. \label{TSP}
\end{align}
The problem \eqref{TSP} is a travelling salesman problem, which can be optimally solved by the one-tree relaxation algorithm via the software Mosek \cite{laporte2}.
In particular, the one-tree relaxation algorithm is an iterative procedure that finds a sequence of one-tree upper bounds to the problem \eqref{TSP} until convergence \cite{laporte2}, and the converged solution is guaranteed to be optimal.

Denoting the optimal solution to the problem \eqref{TSP} as $\{\widehat{\mathbf{W}},\widehat{\lambda}_m\}$,
the optimal objective value of the travelling salesman problem is given by
\begin{align}
&\Upsilon(\widetilde{\mathbf{v}}):=
T-\frac{\mathrm{Tr}(\mathbf{D}^{T}\widehat{\mathbf{W}})}{a}. \label{Upsilon}
\end{align}
Now, by putting the obtained $\Upsilon(\widetilde{\mathbf{v}})$ into $\mathrm{P}3$, the constraint of $\mathrm{P}3$ is written as $\sum_{k=1}^Ks_{k}
= \Upsilon(\widetilde{\mathbf{v}})$.
Assigning a Lagrange multiplier $\rho$ to this constraint, the Lagrangian of $\mathrm{P}3$ is
\begin{align}
\mathcal{L}\left(\{s_{k}\},\rho\right)=&
\mu\left(\alpha_1+\alpha_2a\right)\left[T-\Upsilon(\widetilde{\mathbf{v}})\right]
\nonumber\\
&
+(2-\mu)\sum_{k=1}^K\frac{\gamma_k}{B_k(\widetilde{\mathbf{v}})}
\Theta\left(\frac{s_k}{\gamma_k}\right)
\nonumber\\
&+\rho\left(\sum_{k=1}^Ks_{k}-\Upsilon(\widetilde{\mathbf{v}})\right).
 \nonumber
\end{align}
According to the first-order KKT condition $\frac{\partial \mathcal{L}}{\partial s_k}|_{s_k=\widehat{s}_k}=0$, the optimal $\{\widehat{s}_{k}\}$ and $\widehat{\rho}$ should together satisfy \cite{opt1}:
\begin{align}\label{Theta_gradient}
&-\nabla\Theta\left(\frac{\widehat{s}_k}{\gamma_k}\right)=
\frac{B_k(\widetilde{\mathbf{v}})\widehat{\rho}}{2-\mu},
\end{align}
where $-\nabla\Theta(x)$ is the gradient of $-\Theta(x)$ and is given by
\begin{align}
-\nabla\Theta(x)=1+\mathrm{ln}2\cdot 2^{1/x}/x-2^{1/x},
\end{align}
with $x>0$.
Moreover, as $-\nabla^2\Theta(x)=-
\mathrm{ln}^22\cdot 2^{1/x}/x^3<0$ and
$-\nabla^3\Theta(x)=\mathrm{ln}^32\cdot 2^{1/x}/x^5+3\mathrm{ln}^22\cdot 2^{1/x}/x^4>0$, it can be seen that $-\nabla\Theta(x)$ is a strictly decreasing and convex function of $x$.
As a result, there must exist a strictly decreasing and convex function $\Lambda(x)$ such that $\Lambda(-\nabla\Theta(x))=x$.
That is, the function $\Lambda(x)$ is the inverse function of $-\nabla\Theta(x)$, and it can be numerically computed and stored as a look-up table, with its shape shown in Fig. 3.
Applying $\Lambda(x)$ to both sides of \eqref{Theta_gradient}, we have
\begin{align}\label{sk}
&
\widehat{s}_k=\gamma_k\cdot\Lambda\left(\frac{B_k(\widetilde{\mathbf{v}})\widehat{\rho}}{2-\mu}\right).
\end{align}

\begin{figure}[!t]
\centering
\includegraphics[width=50mm]{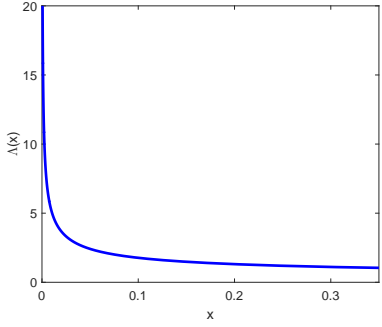}
\caption{The function $\Lambda(x)$.}
\label{fig_sim}
\end{figure}

Notice that in \eqref{sk}, the only unknown is $\widehat{\rho}$, which should satisfy the equality constraint of $\mathrm{P}3$:
\begin{align}
&
\sum_{k=1}^K\gamma_k\cdot\Lambda\left(\frac{B_k(\widetilde{\mathbf{v}})\widehat{\rho}}{2-\mu}\right)
=\Upsilon(\widetilde{\mathbf{v}}).  \label{rho}
\end{align}
Since $\Lambda(x)$ is a decreasing function, bisection search method can be used to find $\widehat{\rho}$ efficiently.
In order to determine the bisection interval, the following proposition (proved in Appendix C) can be established.
\begin{proposition}
The quantity $\widehat{\rho}$ in \eqref{rho} is bounded as
\begin{align}
&-
\frac{(2-\mu)\sum_{k=1}^K\gamma_k}{\sum_{k=1}^K\gamma_kB_k(\widetilde{\mathbf{v}})}\nabla\Theta\left(\frac{\Upsilon(\widetilde{\mathbf{v}})}{\sum_{k=1}^K\gamma_k}\right)
\nonumber\\
&~~~~~~~~~~~~~~~
\leq
\widehat{\rho}
\leq
-\frac{2-\mu}{\mathrm{min}_{l}~B_l(\widetilde{\mathbf{v}})}
~\nabla\Theta\left(\frac{\Upsilon(\widetilde{\mathbf{v}})}{\sum_{k=1}^K\gamma_k}\right). \label{bound}
\end{align}
\end{proposition}
Once $\widehat{\rho}$ is obtained, we can put $\widehat{\rho}$ into \eqref{sk} to get $\{\widehat{s}_{k}\}$.
Further putting $\{\widehat{s}_{k}\}$ into \eqref{qk}, we have
$\widehat{q}_k=\left(2^{\gamma_k/\widehat{s}_k}-1\right)
\Big/B_k(\widetilde{\mathbf{v}})$.

\subsection{Optimal Solution of $\mathbf{v}$}

With path $\widehat{\mathbf{W}}$, transmit times $\{\widehat{s}_{k}\}$, and transmit powers $\{\widehat{q}_{k}\}$ derived in Section IV-B, the optimal objective value of $\mathrm{P}3$ (equivalently $\mathrm{P}2$ with $\mathbf{v}=\widetilde{\mathbf{v}}$) is given by
\begin{align}\label{Xi}
\Xi(\widetilde{\mathbf{v}})=&\mu\left(\alpha_1+\alpha_2a\right)\left[T-\Upsilon\left(\widetilde{\mathbf{v}}\right)\right]
\nonumber\\
&
+(2-\mu)\sum_{k=1}^K\frac{\gamma_k}{B_k\left(\widetilde{\mathbf{v}}\right)}
\Theta\left[\Lambda\left(\frac{B_k\left(\widetilde{\mathbf{v}}\right)
\widehat{\rho}}{2-\mu}\right)\right].
\end{align}
Therefore, problem $\mathrm{P}2$ is re-written as
\begin{align}
\mathrm{P}4:\mathop{\mathrm{min}}_{\substack{\mathbf{v}}}
~~&\Xi(\mathbf{v})
\nonumber\\
~~\mathrm{s.t.}~~&v_{1}=1,~v_{m}\in\{0,1\},~~\forall m=2,\cdots,M. \label{v}
\end{align}
To solve $\mathrm{P}4$, a naive way is to apply exhaustive search for $\mathbf{v}$.
Unfortunately, since the searching space of $\{v_m\}$ is very large (i.e., $2^{M-1}$), direct implementation of exhaustive search is impossible.
To address the above issue, a B$\&$B method is presented for systematically pruning out impossible solutions of $\mathbf{v}$, leading to significant reduction of the computational complexity compared to exhaustive search while guaranteeing the global optimality \cite{bb,MINLP}.

In particular, we define the living pool as a set $\mathcal{Y}$ which stores all the solutions that have not been explored, and the incumbent $I$ as the current best objective value that has been obtained.
Initially, $I$ can be set to the objective value of any feasible solution.
On the other hand, since the feasible set of $\mathbf{v}$ for $\mathrm{P}4$ is $\Omega=\left\{\mathbf{v}\in\{0,1\}^M: v_{1}=1\right\}$,
the initial living pool can be set to $\mathcal{Y}=\{\mathcal{F}_{(1,0)},\mathcal{F}_{(1,1)}\}$ (notice that $\mathcal{Y}$ is a family of sets over $\Omega$), where
$\mathcal{F}_{(1,0)}=\left\{\mathbf{v}\in\Omega: v_{2}=0\right\}$ and $\mathcal{F}_{(1,1)}=\left\{\mathbf{v}\in\Omega: v_{2}=1\right\}$ as shown in Fig. 4a.
It can be seen that $\Omega=\mathcal{F}_{(1,0)}\bigcup\mathcal{F}_{(1,1)}$ and $\mathcal{F}_{(1,0)},\mathcal{F}_{(1,1)}\subset\Omega$.

\begin{figure}[!t]
\centering
\includegraphics[width=70mm]{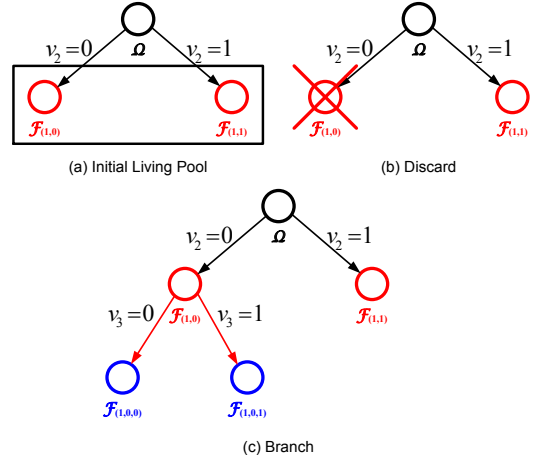}
\caption{Illustration of branch and bound (B\&B) method.}
\label{fig_sim}
\end{figure}

At the beginning, the B$\&$B method computes a lower bound $\Psi(\mathcal{F}_{(1,0)})$ such that $\Psi(\mathcal{F}_{(1,0)})\leq \Xi(\widetilde{\mathbf{v}})$ for any $\widetilde{\mathbf{v}}\in\mathcal{F}_{(1,0)}$, where $\Xi(\widetilde{\mathbf{v}})$ is defined in \eqref{Xi}.
Based on the bounding function value $\Psi(\mathcal{F}_{(1,0)})$, we consider the following three cases.
\begin{itemize}
\item[(i)] If $\Psi(\mathcal{F}_{(1,0)})>I$, then the subset $\mathcal{F}_{(1,0)}$ can be discarded, since no feasible solution inside $\mathcal{F}_{(1,0)}$ leads to better objective than the incumbent. Therefore, we update $\mathcal{Y}\leftarrow\mathcal{Y}\setminus\mathcal{F}_{(1,0)}$. This case is shown in Fig. 4b.

\item[(ii)] If $\Psi(\mathcal{F}_{(1,0)})\leq I$ and $|\mathcal{F}_{(1,0)}|>1$, the possibility of a better solution in $\mathcal{F}_{(1,0)}$ cannot be ruled out. As a result, we need to branch on $\mathcal{F}_{(1,0)}$ and generates two subsets
$\mathcal{F}_{(1,0,0)}=\left\{\mathbf{v}\in\mathcal{F}_{(1,0)}: v_{3}=0\right\}$ and $\mathcal{F}_{(1,0,1)}=\left\{\mathbf{v}\in\mathcal{F}_{(1,0)}: v_{3}=1\right\}$.
By treating $\mathcal{F}_{(1,0,0)}$ and $\mathcal{F}_{(1,0,1)}$ as new subset nodes, we update $\mathcal{Y}\leftarrow\mathcal{Y}\setminus\mathcal{F}_{(1,0)}\cup\{\mathcal{F}_{(1,0,0)},\mathcal{F}_{(1,0,1)}\}$.
This case is shown in Fig. 4c.

\item[(iii)] Otherwise, we must have $|\mathcal{F}_{(1,0)}|=1$. Denoting the unique element in $|\mathcal{F}_{(1,0)}|$ as $\widetilde{\mathbf{v}}$,
we compute $\Xi(\widetilde{\mathbf{v}})$.
If $\Xi(\widetilde{\mathbf{v}})>I$, $\mathcal{F}_{(1,0)}$ is discarded.
On the other hand, if $\Xi(\widetilde{\mathbf{v}})\leq I$, the incumbent is updated as $I\leftarrow\Xi(\widetilde{\mathbf{v}})$ and the current best solution is updated as $\mathbf{v}^\diamond\leftarrow\widetilde{\mathbf{v}}$.
In both cases, since there is no more element in $\mathcal{F}_{(1,0)}$ to be evaluated, we can update $\mathcal{Y}\leftarrow\mathcal{Y}\setminus\mathcal{F}_{(1,0)}$.

\end{itemize}
After $\mathcal{F}_{(1,0)}$ is evaluated, we then evaluate $\mathcal{F}_{(1,1)},\mathcal{F}_{(1,0,0)},\mathcal{F}_{(1,0,1)},\cdots$ by repeating the above procedure until the living pool $\mathcal{Y}$ becomes empty.

For the above B$\&$B method, the key step is to derive the bounding function $\Psi$.
In particular, to guarantee that the proposed B$\&$B method finds the optimal solution of $\mathbf{v}$ to $\mathrm{P}4$,
the bounding function needs to satisfy \cite{bb}:
\begin{align}
&\Psi(\mathcal{F}_{(z_1,\cdots,z_N)})\leq \Xi(\widetilde{\mathbf{v}}),~~\forall~\widetilde{\mathbf{v}}\in\mathcal{F}_{(z_1,\cdots,z_N)},
\label{bounding}
\end{align}
where $(z_1,\cdots,z_N)$ are the values assigned to $(v_1,\cdots,v_N)$ and $N$ is the number of fixed elements in $\mathbf{v}$.
To this end, consider the following bounding function $\Psi$:
\begin{align}
\Psi(\mathcal{F}_{(z_1,\cdots,z_N)})=&\mu\left(\frac{\alpha_1}{a}+\alpha_2\right)\left[T-\Phi\left(\mathbf{z}\right)\right]
\nonumber\\
&
+(2-\mu)\sum_{k=1}^K\frac{\gamma_k}{B_k\left([\mathbf{z}^{T},\mathbf{1}_{M-N}^{T}]^{T}\right)}
\nonumber\\
&
\times
\Theta\left[\Lambda\left(\frac{B_k\left([\mathbf{z}^{T},\mathbf{1}_{M-N}^{T}]^{T}\right)
\delta}{2-\mu}\right)\right], \label{Psi}
\end{align}
where $\mathbf{z}=[z_1,\cdots,z_N]^{T}\in\{0,1\}^N$ (corresponding to the sequence $(z_1,\cdots,z_N)$), and $\delta$ is the solution to
\begin{align}
&\sum_{k=1}^K
\gamma_k\cdot\Lambda\left(\frac{B_k\left([\mathbf{z}^{T},\mathbf{1}_{M-N}^{T}]^{T}\right)
\delta}{2-\mu}\right)=\Phi\left(\mathbf{z}\right).
\label{kappa}
\end{align}
The function
\begin{align}
\Phi\left(\mathbf{z}\right)=&
\mathop{\mathrm{max}}_{\substack{\mathbf{W}}}~\Bigg\{T-\frac{\mathrm{Tr}(\mathbf{D}^{T}\mathbf{W})}{a}:
\nonumber\\
&
\sum_{j=1}^MW_{m,j}=\sum_{j=1}^MW_{j,m}=z_m,
~~\forall m=1,\cdots,N, \nonumber
\\
&W_{m,j}\in\{0,1\},~~\forall m,j,~~W_{m,m}=0,~~\forall m\Bigg\} \label{Phi}
\end{align}
represents a bipartite matching problem, which can be numerically computed via the Hungarian algorithm \cite{hungarian}.
It is proved in Appendix D that \eqref{Psi} satisfies the property \eqref{bounding}.
As a result, by applying the bounding function $\Psi$ in \eqref{Psi}, the proposed B$\&$B method is guaranteed to obtain the optimal solution of $\mathbf{v}$ to problem $\mathrm{P}4$ (equivalently $\mathrm{P}2$).

\begin{algorithm}
    \caption{Computing the optimal solution to $\mathrm{P}1$}
        \begin{algorithmic}[1]
            \State \textbf{Initialize} incumbent $I$ and living pool $\mathcal{Y}=\{\mathcal{F}_{(1,0)},\mathcal{F}_{(1,1)}\}$. Set iteration counter $\mathrm{Iter}=0$.
            \State \textbf{Repeat}
            \State \ \ \ Pick an element $\mathcal{F}_{(z_1,\cdots,z_N)}$ from the living pool $\mathcal{Y}$, and compute the bounding function $\Psi(\mathcal{F}_{(z_1,\cdots,z_N)})$.
            \State \ \ \ If $\Psi(\mathcal{F}_{(z_1,\cdots,z_N)})\leq I$
            \State \ \ \ \ \ \ If $|\mathcal{F}_{(z_1,\cdots,z_N)}|>1$
            \State \ \ \ \ \ \ \ \ \ Branch on $\mathcal{F}_{(z_1,\cdots,z_N)}$ generating $\mathcal{F}_{(z_1,\cdots,z_N,0)}$ and $\mathcal{F}_{(z_1,\cdots,z_N,1)}$.
            \State \ \ \ \ \ \ \ \ \ Update $\mathcal{Y}\leftarrow\mathcal{Y}\cup
            \{\mathcal{F}_{(z_1,\cdots,z_N,0)},\mathcal{F}_{(z_1,\cdots,z_N,1)}\}$.
            \State \ \ \ \ \ \ Else
            \State \ \ \ \ \ \ \ \ \  Compute $\Xi(\widetilde{\mathbf{v}})$ with $\widetilde{\mathbf{v}}\in\mathcal{F}_{(z_1,\cdots,z_N)}$.
            \State \ \ \ \ \ \ \ \ \  If $\Xi(\widetilde{\mathbf{v}})\leq I$
            \State \ \ \ \ \ \ \ \ \ \ \ \ Update $I\leftarrow\Xi(\widetilde{\mathbf{v}})$ and $\mathbf{v}^\diamond\leftarrow\widetilde{\mathbf{v}}$.
            \State \ \ \ \ \ \ \ \ \ End
            \State \ \ \ \ \ \ End
            \State \ \ \ End
            \State \ \ \ Update $\mathcal{Y}\leftarrow\mathcal{Y}\setminus\mathcal{F}_{(z_1,\cdots,z_N)}$.
            \State \ \ \ $\mathrm{Iter}\leftarrow \mathrm{Iter}+1$.
            \State \textbf{Until} $\mathcal{Y}=\emptyset$ and the optimal $\mathbf{v}^*=\mathbf{v}^\diamond$.
            \State Compute the optimal $\mathbf{W}^*$ using \eqref{TSP}.
            \State Compute the optimal $t_{k,m}^*$ using \eqref{tkm}, where $s_k^*$ is given by \eqref{sk} and $\rho^*$ is given by \eqref{rho}.
            \State Compute the optimal $p_{k,m}^*$ using \eqref{pkm}, where $q_k^*$ is given by \eqref{qk}.
            \State Output $\{\mathbf{v}^*,\mathbf{W}^*,t_{k,m}^*,p_{k,m}^*\}$.
        \end{algorithmic}
\end{algorithm}

\subsection{Summary of Algorithm and Complexity Analysis}

Since the B$\&$B algorithm finds the optimal solution of $\mathbf{v}$ to $\mathrm{P}2$, and the optimal solution of $\mathbf{W}$ and $\{s_{k},q_{k}\}$ with fixed $\mathbf{v}$ can be computed according to Section IV-B, the entire algorithm for computing the optimal solution to $\mathrm{P}2$ (equivalently $\mathrm{P}1$) is summarized in Algorithm 1.

In terms of computational complexity, computing $\Phi\left(\mathbf{z}\right)$ in \eqref{Phi} via the Hungarian algorithm requires a complexity of $O\left((2M)^3\right)$ \cite{hungarian}.
On the other hand, computing $\delta$ would require bisection search in solving \eqref{kappa} and the number of iterations is given by $\mathrm{log}_2\left(\frac{C}{\epsilon}\right)$ \cite{bisection}, where $C$ is the length of the initial searching interval given by \eqref{bound} and $\epsilon$ is the target accuracy.
Therefore, computing the bounding function $\Psi$ requires a complexity of $O\left((2M)^3+K\mathrm{log}_2\left(\frac{C}{\epsilon}\right)\right)$.
Finally, computing $\Xi\left(\widetilde{\mathbf{v}}\right)$ for a fixed $\mathbf{v}=\widetilde{\mathbf{v}}$ would involve the travelling salesman problem, which requires a complexity of $O\left((M-1)^2\cdot2^{M-1}\right)$ in the worst case \cite{tsp}.

Based on the above analysis, the total complexity of Algorithm 1 is given by
\begin{align}
\mathrm{Comp}=&O\Big[X_1\left((2M)^3
+K\mathrm{log}_2\left(\frac{C}{\epsilon}\right)\right)
\nonumber\\
&
+X_2(M-1)^2\cdot2^{M-1}\Big],
\end{align}
where $X_1$ is the number of times for computing $\Psi$ and $X_2$ is the number of times for computing $\Xi\left(\widetilde{\mathbf{v}}\right)$.
In contrast, the computational complexity for exhaustive search of $\mathbf{v}$ is $2^{M-1}\cdot(M-1)^2\cdot2^{M-1}$.
As $(2M)^3+K\mathrm{log}_2(C/\epsilon)$ is much smaller than $(M-1)^2\cdot 2^{M-1}$, and by simulation $X_1+X_2$ is significantly smaller than $2^{M-1}$, the proposed Algorithm 1 can significantly reduce the computational complexity compared to exhaustive search.

Finally, for the above algorithm, we need an initial incumbent of $I$.
Theoretically, $I$ can be set to the function value of any feasible solution to $\mathrm{P}1$.
However, since we are minimizing the objective function of $\mathrm{P}1$, a smaller initial incumbent $I$ could help in reducing the size of the living pool \cite{hybrid}, and the next section will derive an efficient initialization method.

\emph{Remark 3:} If the transmit power is limited due to hardware constraints (e.g., cost of an amplifier), we can add a constraint $s_{k}\leq P_{\mathrm{max}}$ (with $P_{\mathrm{max}}$ being the upper limit of transmit power) for all $k$ to problem $\mathrm{P}3$.
In such a case, the proposed Algorithm 1 is still applicable if we modify the objective function $\Xi(\mathbf{v})$ in problem $\mathrm{P}4$ into
\begin{align}\label{tildeXi}
\widetilde{\Xi}(\mathbf{v})=&\mathop{\mathrm{min}}_{\substack{\{s_{k}\}}}
~\Bigg\{\mu\left(\alpha_1+\alpha_2a\right)\left(T-\Upsilon\left(\mathbf{v}\right)\right)
\nonumber\\
&
+(2-\mu)\sum_{k=1}^K\frac{\gamma_k}{B_k(\mathbf{v})}
\Theta\left(\frac{s_k}{\gamma_k}\right):
\nonumber\\
&\sum_{k=1}^Ks_{k}
=\Upsilon\left(\mathbf{v}\right),~~0<s_{k}\leq P_{\mathrm{max}},~~\forall k
\Bigg\},
\end{align}
and the bounding function $\Psi$ in \eqref{Psi} into
\begin{align}\label{tildePsi}
&\widetilde{\Psi}(\mathcal{F}_{(z_1,\cdots,z_N)})=\mathop{\mathrm{min}}_{\substack{\{s_{k}\}}}
~\Bigg\{\mu\left(\alpha_1+\alpha_2a\right)\left(T-\Phi\left(\mathbf{z}\right)\right)
\nonumber\\
&
+(2-\mu)\sum_{k=1}^K\frac{\gamma_k}{B_k\left([\mathbf{z}^T,\mathbf{1}_{M-N}^T]^T\right)}
\Theta\left(\frac{s_k}{\gamma_k}\right):
\nonumber\\
&\sum_{k=1}^Ks_{k}
=\Phi\left(\mathbf{z}\right),~~0<s_{k}\leq P_{\mathrm{max}},~~\forall k
\Bigg\}.
\end{align}

\section{Initialization via Successive Local Search}

In order to obtain a good initial incumbent of $I$, this section proposes a local optimal solution method based on successive local search \cite{meta,rls}.
More specifically, we start from a feasible solution of $\mathbf{v}$ (e.g., $\mathbf{v}^{[0]}=[1,0,\cdots,0]^{T}$), and randomly selects a candidate solution $\mathbf{v}'$ from the neighborhood $\mathcal{N}(\mathbf{v}^{[0]})$. Since a natural neighborhood operator for binary optimization is to flip the value of $\{v_m\}$, $\mathcal{N}(\mathbf{v}^{[0]})$ can be set to
\begin{align}
\mathcal{N}(\mathbf{v}^{[0]})=\{\mathbf{v}:||\mathbf{v}-\mathbf{v}^{[0]}||_0\leq L,~\mathbf{v}\in\Omega\},
\end{align}
where $L\geq 1$ is the size of neighborhood.
It can be seen that $\mathcal{N}(\mathbf{v}^{[0]})$ is a subset of the entire feasible space $\Omega$ and containing solutions ``close'' to $\mathbf{v}^{[0]}$.

With the neighborhood $\mathcal{N}(\mathbf{v}^{[0]})$ defined above and the choice of $\mathbf{v}$ fixed to $\mathbf{v}=\mathbf{v}'$, we consider two cases.
\begin{itemize}
\item[(i)] If $\Xi(\mathbf{v}')\leq\Xi(\mathbf{v}^{[0]})$, we update $\mathbf{v}^{[1]}\leftarrow\mathbf{v}'$. By treating $\mathbf{v}^{[1]}$ as a new feasible solution, we can construct the next neighborhood $\mathcal{N}(\mathbf{v}^{[1]})$.

\item[(ii)] If $\Xi(\mathbf{v}')>\Xi(\mathbf{v}^{[0]})$, we find another point within the neighborhood $\mathcal{N}(\mathbf{v}^{[0]})$ until $\Xi(\mathbf{v}')\leq\Xi(\mathbf{v}^{[0]})$.

\end{itemize}
The above procedure is repeated to generate a sequence of $\{\mathbf{v}^{[1]},\mathbf{v}^{[2]},\cdots\}$ and the converged point is guaranteed to be a local optimal solution to $\mathrm{P}1$ \cite{dimitri}.
But since our aim is to obtain a good initial incumbent, it is not necessary to wait until the successive local search converges.
In fact, we can terminate the iterative procedure when the number of iterations is larger than $\overline{\mathrm{Iter}}$.
Denoting the solution after $\overline{\mathrm{Iter}}$ iterations as $\mathbf{v}^{\diamond}$, we can set the initial incumbent as $I=\Xi(\mathbf{v}^\diamond)$.
The entire procedure to generate an initial incumbent is summarized in Algorithm 2, and the complexity of executing Algorithm 2 is $\overline{\mathrm{Iter}}\cdot(M-1)^2\cdot2^{M-1}$.

\begin{algorithm}
    \caption{Initialization via successive local search}
        \begin{algorithmic}[1]
            \State \textbf{Initialize} $\mathbf{v}^{[0]}=[1,0,\cdots,0]^{T}$ and $L=3$. Set $n=0$ and iteration counter $\mathrm{Iter}=0$.
            \State \textbf{Repeat}
            \State \ \ \ Sample a solution $\mathbf{v}'\in\mathcal{N}(\mathbf{v}^{[n]})$.
            \State \ \ \ Compute $\Xi(\mathbf{v}')$ using \eqref{Xi}.
            \State \ \ \ If $\Xi(\mathbf{v}')\leq\Xi(\mathbf{v}^{[n]})$, update $\mathbf{v}^{[n+1]}\leftarrow\mathbf{v}'$ and $n\leftarrow n+1$.
            \State \ \ \ Update $\mathrm{Iter}\leftarrow \mathrm{Iter}+1$.
            \State \textbf{Until} $\mathrm{Iter}=\overline{\mathrm{Iter}}$.
            \State Output initial incumbent $I=\Xi(\mathbf{v}^{[n]})$.
        \end{algorithmic}
\end{algorithm}

\section{Simulation Results and Discussions}

This section provides simulation results to evaluate the performance of the UGV backscatter communication system.
It is assumed that the backscattering efficiency is $\eta=0.78$ (corresponding to $1.1~\mathrm{dB}$ loss \cite{back11}), the performance loss due to imperfect modulation is $\beta=0.5$ \cite{back11}, and the weighting factor $\mu=1$.
Within the time budget $T=500~\mathrm{s}$, the data collection targets $\gamma_k\sim \mathcal{U}(1,2)$ in the unit of $\mathrm{bit/Hz}$ are requested by $K=50$ IoT users (corresponding to a spectral efficiency of $K\gamma_k/T=0.1\sim0.2~\mathrm{bps/Hz}$ \cite{iot1}), where $\mathcal{U}(a,b)$ is the uniform distribution within the interval $[a,b]$.

Based on the above settings, we simulate the data collection map as a $20~\mathrm{m}\times20~\mathrm{m}=400~\mathrm{m}^2$ square area, which is a typical size for smart warehouses.
Inside this map, $K=50$ IoT users\footnote{Notice that $K$ is the number of IoT users assigned to the considered UGV. There might exist other IoT users, which can be inactive or assigned to other UGVs.} and $M=12$ vertices representing stopping points are uniformly scattered.
Among all the vertices, the vertex $m=1$ is selected as the starting point of the UGV.
With the locations of all the stopping points and the IoT devices, the distances between each IoT device and stopping point can be computed, and the distance-dependent path-loss model $\varrho_{k,m}=\varrho_0\cdot(\frac{d_{k,m}}{d_0})^{-2.5}$ is adopted \cite{channel}, where $d_{k,m}$ is the distance from user $k$ to the stopping point $m$, and $\varrho_0=10^{-3}$ is the path-loss at $d_0=1~\mathrm{m}$.
Based on the path-loss model, channels $g_{k,m}$ and $h_{k,m}$ are generated according to $\mathcal{CN}(0,\varrho_{k,m})$.
Each point in the figures is obtained by averaging over $100$ simulation runs, with independent channels and realizations of locations of vertices and users in each run.

\begin{figure}[!t]
\centering
\includegraphics[width=70mm]{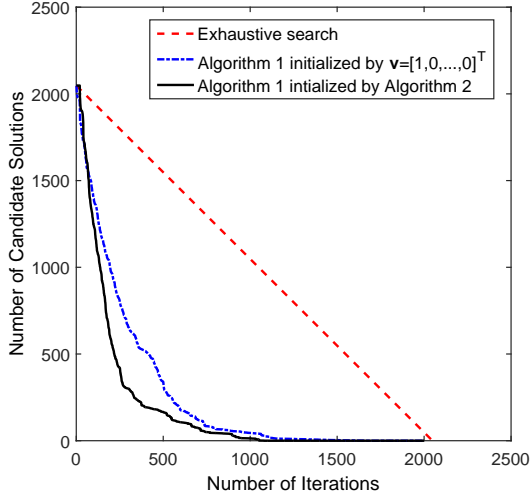}
\caption{Number of candidate solutions versus the number of iterations for the case of $K=50$ and $M=12$ at $N_0=-95~\mathrm{dBm}$.}
\label{fig_sim}
\end{figure}

In order to verify the effectiveness of Algorithm 1 in Section IV, Fig. 5 shows the number of candidate solutions in the living pool $\mathcal{Y}$ versus the number of iterations (represented by $\mathrm{Iter}$ in Algorithm 1) when the receiver noise power $N_0=-95~\mathrm{dBm}$  (corresponding to power spectral density $-145~\mathrm{dBm/Hz}$ with $100~\mathrm{kHz}$ bandwidth \cite{iot1}).
It can be seen that even with a naive initial incumbent obtained by setting $\mathbf{v}=[1,0,\cdots,0]^{T}$, the proposed Algorithm 1 still leads to a significantly faster decrease in the number of candidate solutions than the exhaustive search.
Moreover, by using Algorithm 2 to provide an initialization, the number of iterations for Algorithm 1 (together with $20$ iterations of Algorithm 2) to reach zero candidate solution can be further decreased.
Notice that in each iteration,  Algorithm 1 requires a smaller computational complexity than that of the exhaustive search according to Section IV-C.
Therefore, the total complexity of Algorithm 1 is significantly reduced compared to the exhaustive search.

\begin{figure}
 \centering
  \subfigure[]{
    \label{fig:subfig:b} 
    \includegraphics[width=70mm]{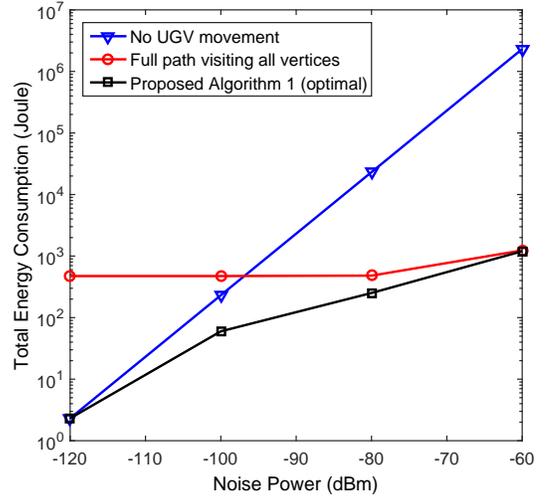}}
          \hspace{0.02in}
      \subfigure[]{
    \label{fig:subfig:b} 
    \includegraphics[width=70mm]{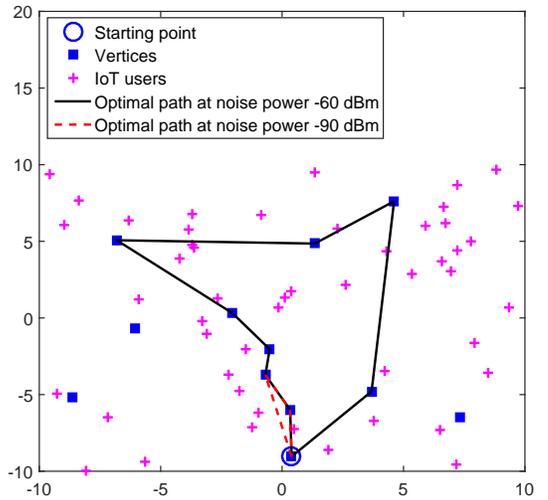}}
  \caption{(a) Total energy consumption versus noise power $N_0$ with $K=50$ and $M=12$ when $\gamma_k\sim \mathcal{U}(1,2)$; (b) The optimal path with $K=50$ and $M=12$ at noise power $N_0=-60~\mathrm{dBm}$ and  $N_0=-90~\mathrm{dBm}$.
}
  \label{fig:subfig} 
\end{figure}

Next, we focus on the energy management performance of Algorithm 1.
In particular, the case of $K=50$ with $M=12$ is simulated, and the total energy consumption versus the noise power $N_0$ is shown in Fig. 6a.
For comparison, we also simulate the scheme with no UGV movement (i.e., the optimal solution to $\mathrm{P}1$ with $\mathbf{v}=[1,0,\cdots,0]^T$) and the scheme with full path visiting all vertices (i.e., the optimal solution to $\mathrm{P}1$ with $\mathbf{v}=\mathbf{1}$).
It can be seen that if the noise power is large, by allowing the UGV to visit all the vertices, it is possible to achieve a significantly lower energy consumption compared to the case of no UGV movement.
However, this conclusion does not hold in the small noise power regime, which indicates that moving is not always beneficial.
Fortunately, the proposed Algorithm 1 can automatically determine whether to move and how far to move.
For example, if the noise power is extremely small (e.g., $-120~\mathrm{dBm}$), the UGV could easily collect the data from IoT users at the starting point.
In such a case, the proposed Algorithm 1 would fix the UGV at the starting point.
This can be seen from Fig. 6a at $N_0=-120~\mathrm{dBm}$, in which Algorithm 1 leads to the same performance as the scheme of no UGV movement.
However, if the noise power is increased to a medium value (e.g., $-90$ $\mathrm{dBm}$), the total energy is reduced by allowing the UGV to move (with the moving path being the red line shown in Fig. 6b).
On the other hand, if the noise power is large (e.g., $-60$ $\mathrm{dBm}$), the energy for data collection would be high for far-away users.
Therefore, the UGV should spend more motion energy to get closer to IoT users.
This is the black line shown in Fig. 6b.
But no matter which case happens, the proposed algorithm adaptively finds the best trade-off between spending energy on moving versus on communication, and therefore achieves the minimum energy consumption for all the simulated values of $N_0$ as shown in Fig. 6a.
Notice that the largest transmit power for the proposed Algorithm 1 in Fig. 6 is $30.8~\mathrm{W}$ (occurred at noise power $-60~\mathrm{dBm}$).
Translating this number to the received power density at $1~\mathrm{m}$ gives $2.45~\mathrm{W/m^2}$, which is within the requirement ($<10~\mathrm{W/m^2}$) set by the IEEE standard C95.1-2005 \cite[Remark 4]{wpt}.

The above Fig. 6 has shown that the noise power $N_0$ can affect the path obtained from Algorithm 1.
In fact, other parameters such as the time budget $T$ and backscatter efficiency $\eta$ could also impact the path.
To see this, the case of $K=50$ with $M=12$ at noise power $N_0=-80~\mathrm{dBm}$ is simulated, and the paths for
$(T,\eta)=(500~\mathrm{s},0.78)$, $(T,\eta)=(20~\mathrm{s},0.78)$, and $(T,\eta)=(500~\mathrm{s},0.1)$ are compared in Fig. 7.
It can be seen from Fig. 7 that the path for $(T,\eta)=(20~\mathrm{s},0.78)$ involves a smaller moving distance than that for $(T,\eta)=(500~\mathrm{s},0.78)$.
This is because a smaller $T$ would restrict the constraint \eqref{P1b} of $\mathrm{P}1$, which forces the UGV to reduce its motion time and moving distance.
On the other hand, the path for $(T,\eta)=(500~\mathrm{s},0.1)$ involves a larger moving distance than that for $(T,\eta)=(500~\mathrm{s},0.78)$, since a smaller $\eta$ would deteriorate the communication qualities, which forces the UGV to get closer to IoT users.

In order to evaluate the performance of the proposed algorithm under various QoS requirements, the total energy consumption versus the data collection target $\gamma_1=\gamma_2=\cdots=\gamma_K$ in $\mathrm{bit/Hz}$ at noise power $N_0=-90$ $\mathrm{dBm}$, is shown in Fig. 8.
It can be seen that under all the simulated values of data collection targets, the proposed Algorithm 1 always achieves the minimum energy consumption.
Moreover, the blue and black curves intersect on the left hand side, while the red and black curves intersect on the right side.
This means that for a very low data collection target, the proposed algorithm results in a non-moving UGV, and for a very high data collection target,
the UGV would visit all the vertices.
Notice that the red curve in Fig. 6 and Fig. 8 is in fact increasing slowly.
The reason behind such a slow change is that the communication energy is negligible compared to the energy required for mobility if the UGV visits all vertices.

\begin{figure}[!t]
\centering
\includegraphics[width=70mm]{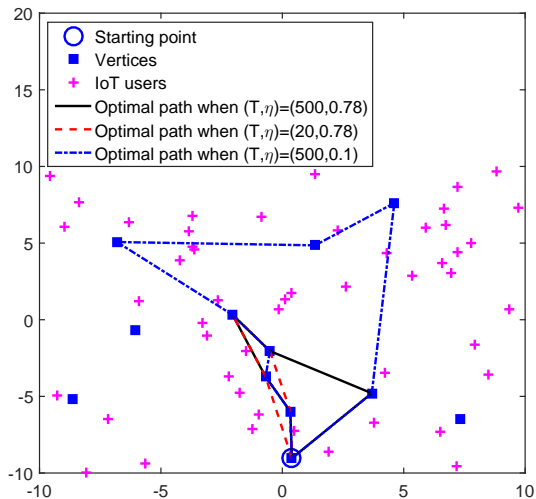}
\caption{The optimal paths with $K=50$ and $M=12$ at $N_0=-80~\mathrm{dBm}$ for different values of $(T,\eta)$.}
\label{fig_sim}
\end{figure}

\begin{figure}[!t]
\centering
\includegraphics[width=70mm]{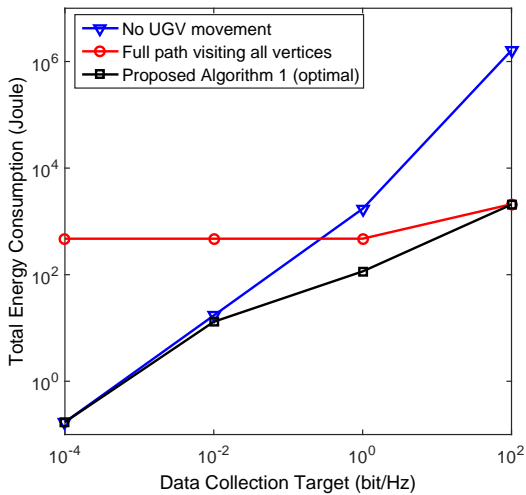}
\caption{Total energy consumption versus data collection target $\gamma_k$ with $K=50$ and $M=12$ at $N_0=-90$ $\mathrm{dBm}$.}
\label{fig_sim}
\end{figure}

To further assess the performance of the proposed Algorithm 1 when $\mu$ varies, the case of $K=50$ with $M=12$ at $N_0=-90~\mathrm{dBm}$ is simulated and the result is shown in Fig. 9.
It can be seen that if $\mu=0$, the proposed Algorithm 1 has the same performance as that of the full path.
This is because $\mu=0$ would make the motion energy disappear in the objective function of $\mathrm{P}1$, and the best strategy is to allow the UGV to visit all vertices.
However, even with a slight increase in $\mu$, the proposed Algorithm 1 would outperform other benchmark schemes.
Finally, it can be seen that if $\mu$ increases, the motion energy decreases but the communication energy increases.
Therefore, $\mu$ can be used to adjust the relative amount of communication energy versus motion energy.
Notice that the communication energy of UGV in Fig. 9 is in the same order of magnitude as the motion energy of UGV when $\mu=1$.
This is different from UAV communications, where the propulsion energy of UAV is much larger than the communication energy to keep the UAV aloft \cite{uav4,uav5}.

\begin{figure}[!t]
\centering
\includegraphics[width=70mm]{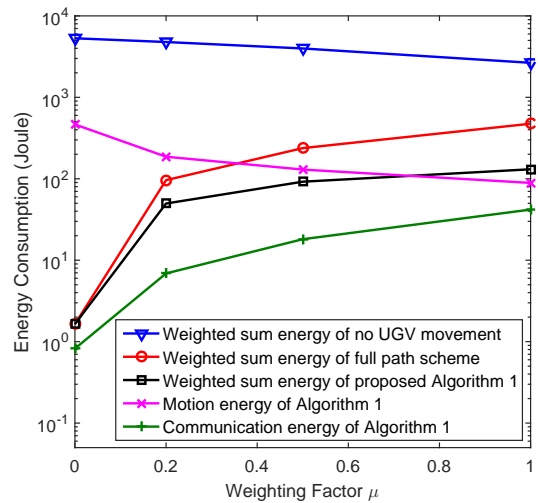}
\caption{Energy consumption versus $\mu$ for the case of $K=50$ with $M=12$ at $N_0=-90~\mathrm{dBm}$.}
\label{fig_sim}
\end{figure}

\begin{figure}[!t]
\centering
\includegraphics[width=70mm]{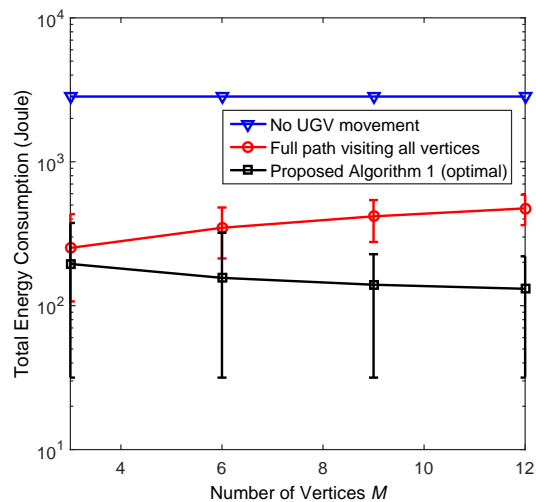}
\caption{Total energy consumption versus $M$ for the case of $K=50$ at $N_0=-90$ $\mathrm{dBm}$. The vertical lines indicate the best case and the worst case performance.
But since the performance with no UGV movement (blue line) is much worse than the other two schemes, the uncertainty bars on the blue line are not plotted.
}
\label{fig_sim}
\end{figure}

Finally, we analyze the impact of the number of vertices $M$ on the energy consumption.
In particular, the case of $K=50$ with $M\in\{3,6,9,12\}$ at noise power $N_0=-90~\mathrm{dBm}$ is simulated, and the result is shown in Fig. 10.
It can be seen from Fig. 10 that the scheme with no UGV movement is independent of $M$, since the UGV is fixed at the starting point and would not visit any other vertex.
On the other hand, the performance of the full path visiting all vertices becomes worse when $M$ increases.
This is because the UGV needs to visit more vertices and therefore consumes more motion energy.
In addition, the total energy consumption of the proposed Algorithm 1 decreases when $M$ increases, since a larger $M$ would give the algorithm more freedom to optimize the trajectory.
Lastly, for a fixed value of $M$, the performance gap between the best case and the worst case could be large as shown in the error bars of Fig. 10.
This is due to the different spatial distributions of users in different map realizations.
If the users are distributed in clusters (e.g., users' locations follow Gaussian mixture distribution), the proposed algorithm could achieve the best performance.
On the other hand, if the users are widely spread out in the map, it would be difficult to collect data from these users, resulting in the worst case performance of the algorithm.
It is worth noting that the locations of vertices are in general independent of the users' locations.
This is because the vertices should be placed where the UGV is able to approach and stop.
However, if the locations of vertices are allowed to be chosen freely, a promising heuristic for setting the locations of vertices is to cluster the $K$ users into $M$ groups and place the vertices at the cluster centers \cite{cluster}.

\section{Conclusions}

This paper studied a UGV-based backscatter data collection system, with an integrated graph mobility model and backscatter communication model.
The joint mobility management and power allocation problem was formulated with the aim of energy minimization subject to communication QoS constraints and mobility graph structure constraints.
An algorithm that achieves the optimal solution was derived, and it automatically balances the trade-off between spending energy on moving and on communication.
Simulation results showed that the proposed algorithm could significantly save energy consumption compared to the scheme with no UGV movement and the scheme with a fixed moving path.

\appendices

\section{Proof of Proposition 1}

This proposition contains two parts, and we will first prove part (ii) by contradiction.

\subsection{Proof of Part (ii)}

To prove part (ii) by contradiction, consider an optimal solution $\{t_{k,m}^*,p_{k,m}^*,\mathbf{v}^*,\mathbf{W}^*,\lambda_m^*\}$ to $\mathrm{P}1$ with a particular
$(i,j)$ such that $t_{i,j}^*=\widetilde{t}\neq 0$.
Assume that the corresponding $p_{i,j}^*=0$.

Since $\{t_{k,m}^*,p_{k,m}^*,\mathbf{v}^*,\mathbf{W}^*,\lambda_m^*\}$ is optimal to $\mathrm{P}1$, it must satisfy \eqref{P1a} of $\mathrm{P}1$, i.e., there must exist some $n\neq j$ such that $t_{i,n}^*=t'\neq0$ and $p_{i,n}^*=p'\neq0$.
Now, consider a related problem of $\mathrm{P}1$ by fixing all the variables to their optimal values except for $(t_{i,j},t_{i,n},p_{i,j},p_{i,n})$:
\begin{align}
\mathop{\mathrm{min}}_{\substack{t_{i,j},t_{i,n},p_{i,j},p_{i,n}\geq 0}}
~~&t_{i,j}p_{i,j}+t_{i,n}p_{i,n}
\nonumber\\
~~~~~~\mathrm{s.t.}~~~~~~~~~~&C_1+t_{i,j}\mathrm{log}_2\left(1+v_j^*A_{i,j}p_{i,j}\right)
\nonumber\\
&
+t_{i,n}\mathrm{log}_2\left(1+v_n^*A_{i,n}p_{i,n}\right)\geq \gamma_i, \nonumber
\\
&C_2+t_{i,j}+t_{i,n}\leq T, \label{Appendix A 1}
\end{align}
where
\begin{align}
C_1=&\sum_{(k,m)\notin\{(i,j),(i,n)\}}t_{k,m}^*\mathrm{log}_2\left(1+v_m^*A_{k,m}p_{k,m}^*\right),
\nonumber\\
C_2=&\sum_{(k,m)\notin\{(i,j),(i,n)\}}t_{k,m}^*+\frac{\mathrm{Tr}(\mathbf{D}^{T}\mathbf{W}^*)}{a}. \nonumber
\end{align}
As $\{t_{k,m}^*,p_{k,m}^*,\mathbf{v}^*,\mathbf{W}^*\}$ is optimal to $\mathrm{P}1$, it can be seen that
\begin{align}
&(t_{i,j},t_{i,n},p_{i,j},p_{i,n})=(\widetilde{t},t',0,p')~\mathrm{is~optimal~to}~\eqref{Appendix A 1}. \label{Appendix A 2}
\end{align}
Therefore, $(\widetilde{t},t',0,p')$ should satisfy the constraints of \eqref{Appendix A 1}, which leads to
\begin{align}
&C_1+t'\mathrm{log}_2\left(1+v_n^*A_{i,n}p'\right)\geq \gamma_i,~~
C_2+\widetilde{t}+t'\leq T. \label{Appendix A 3}
\end{align}
Furthermore, by Jensen's inequality, we have
\begin{align}
&(\widetilde{t}+t')\mathrm{log}_2\left(1+v_n^*A_{i,n}\cdot\frac{\widetilde{t}\cdot0+t'p'}{\widetilde{t}+t'}\right)
\nonumber\\
&
>
\widetilde{t}\mathrm{log}_2\left(1+v_n^*A_{i,n}\cdot 0\right)+
t'\mathrm{log}_2\left(1+v_n^*A_{i,n}p'\right), \label{Appendix A 4}
\end{align}
where the strict inequality is due to $\widetilde{t},t',p' \neq 0$.
Adding $C_1$ to both sides of \eqref{Appendix A 4}, and combining the first inequality of \eqref{Appendix A 3}, we have $C_1+(\widetilde{t}+t')\mathrm{log}_2\left(1+v_n^*A_{i,n}\cdot\frac{t'}{\widetilde{t}+t'}p'\right)>\gamma_i$.
Comparing this result and the second inequality of \eqref{Appendix A 3} to the constraints of \eqref{Appendix A 1}, it can be seen that $(t_{i,j},t_{i,n},p_{i,j},p_{i,n})=(0,\widetilde{t}+t',0,\frac{t'}{\widetilde{t}+t'}p'-\Delta p)$ is feasible for \eqref{Appendix A 1} under sufficiently small $\Delta p>0$.
Putting $(0,\widetilde{t}+t',0,\frac{t'}{\widetilde{t}+t'}p'-\Delta p)$ and $(\widetilde{t},t',0,p')$ from \eqref{Appendix A 2} into the objective function of \eqref{Appendix A 1}, we obtain $t'p'-(\widetilde{t}+t')\Delta p$ and $t'p'$, respectively.
Further due to $t'p'-(\widetilde{t}+t')\Delta p<t'p'$, it is clear that $(\widetilde{t},t',0,p')$ cannot be optimal to \eqref{Appendix A 1}.
This contradicts to \eqref{Appendix A 2}.
Therefore, $p_{i,j}^*\neq 0$.

\subsection{Proof of the First Part of (i)}

The first part of (i) can be proved by following a similar procedure to that of part (ii).
In particular, consider an optimal solution $\{t_{k,m}^*,p_{k,m}^*,\mathbf{v}^*,\mathbf{W}^*,\lambda_m^*\}$ to $\mathrm{P}1$.
Assume that there exists some user $i$ such that $t_{i,j}^*=0$ at vertex $j=\mathop{\mathrm{argmax}}_{l\in\mathcal{V}}~v_l^*A_{i,l}$.

Since $\{t_{k,m}^*,p_{k,m}^*,\mathbf{v}^*,\mathbf{W}^*,\lambda_k^*\}$ is optimal to $\mathrm{P}1$, it must satisfy \eqref{P1a} of $\mathrm{P}1$, i.e., there must exist some $n\neq j$ such that $t_{i,n}^*=t'\neq0$ and $p_{i,n}^*=p'\neq0$.
Furthermore, under $\{t_{k,m}^*,p_{k,m}^*,\mathbf{v}^*,\mathbf{W}^*,\lambda_m^*\}$ with $t_{i,j}^*=0$, it can be shown that $(t_{i,j},t_{i,n},p_{i,j},p_{i,n})=(0,t',p_{i,j}^*,p')$ is optimal to \eqref{Appendix A 1}.
As a result, $(0,t',p_{i,j}^*,p')$ must satisfy the constraints of \eqref{Appendix A 1}, i.e.,
\begin{align}
&C_1+t'\mathrm{log}_2\left(1+v_n^*A_{i,n}p'\right)\geq \gamma_i,~~
C_2+t'\leq T,
\end{align}
which can be re-written as
\begin{align}
&C_1+t'\mathrm{log}_2\left(1+v_j^*A_{i,j}\cdot\frac{v_n^*A_{i,n}}{v_j^*A_{i,j}}p'\right)\geq \gamma_i,~~C_2+t'\leq T. \label{Appendix A 5}
\end{align}
Comparing \eqref{Appendix A 5} with the constraints of \eqref{Appendix A 1}, it can be seen that $(t_{i,j},t_{i,n},p_{i,j},p_{i,n})=(t',0,\frac{v_n^*A_{i,n}}{v_j^*A_{i,j}}p',0)$ is also feasible for \eqref{Appendix A 1}.
Putting $(0,t',p_{i,j}^*,p')$ and $(t',0,\frac{v_n^*A_{i,n}}{v_j^*A_{i,j}}p',0)$ into the objective function of \eqref{Appendix A 1}, we obtain $p't'$ and
$\frac{v_n^*A_{i,n}}{v_j^*A_{i,j}}p't'$, respectively.
Further due to $(0,t',p_{i,j}^*,p')$ being optimal to \eqref{Appendix A 1}, we have $p't'\leq \frac{v_n^*A_{i,n}}{v_j^*A_{i,j}}p't'$, which leads to
$v_j^*A_{i,j}\leq v_n^*A_{i,n}$.
As $A_{i,j}$ and $A_{i,n}$ are not equal almost surely, and $v_j^*,v_n^* \in \{0,1\}$, the equality sign in $v_j^*A_{i,j}\leq v_n^*A_{i,n}$ cannot hold.
This gives us $v_j^*A_{i,j}<v_n^*A_{i,n}$, but it contradicts to $j=\mathop{\mathrm{argmax}}_{l\in\mathcal{V}}~v_l^*A_{i,l}$.
Therefore, $t_{i,j}^*\neq 0$ at vertex $j=\mathop{\mathrm{argmax}}_{l\in\mathcal{V}}~v_l^*A_{i,l}$.

\subsection{Proof of the Second Part of (i)}

We will prove the second part of (i) by contradiction.
In particular, assume that there exists some user $i$ such that $t^*_{i,n}\neq 0$ at vertex $n\neq \mathop{\mathrm{argmax}}_{l\in\mathcal{V}}~v_l^*A_{i,l}$.
On the other hand, based on the first part of (i) of \textbf{Proposition 1}, we also have $t^*_{i,j}\neq 0$ at vertex $j=\mathop{\mathrm{argmax}}_{l\in\mathcal{V}}~v_l^*A_{i,l}$.
Correspondingly, by part (ii) of \textbf{Proposition 1}, we have $p^*_{i,n},p^*_{i,j}\neq 0$.

Now, the partial Lagrangian of $\mathrm{P}1$ with respect to $\{t_{k,m},p_{k,m}\}$ under fixed $\mathbf{v}=\mathbf{v}^*$ is
\begin{align}
&\mathcal{L}\left(\{t_{k,m},p_{k,m}\},\{\zeta_{k},\varphi,\chi_{k,m},\theta_{k,m},\xi_{k,m}\}\right)
\nonumber\\
=&(2-\mu)\sum_{m=1}^M\sum_{k=1}^Kt_{k,m}p_{k,m}
\nonumber\\
&
+\sum_{k=1}^K
\zeta_k\left[\gamma_k-\frac{1}{T}\sum_{m=1}^M
t_{k,m}\mathrm{log}_2\left(1+v_m^*A_{k,m}p_{k,m}\right)\right]
\nonumber\\
&
+\varphi\left(\sum_{m=1}^M\sum_{k=1}^Kt_{k,m}+
\frac{1}{a}\mathrm{Tr}(\mathbf{D}^{T}\mathbf{W})-T\right)
\nonumber\\
&
+\sum_{m=1}^M\sum_{k=1}^K\chi_{k,m}\left[(1-v_m^*)t_{k,m}\right]
\nonumber\\
&-\sum_{m=1}^M\sum_{k=1}^K\theta_{k,m}t_{k,m}-\sum_{m=1}^M\sum_{k=1}^K\xi_{k,m}p_{k,m}
,
 \nonumber
\end{align}
where $\{\zeta_{k},\varphi,\chi_{k,m},\theta_{k,m},\xi_{k,m}\}$ are Lagrange multipliers.
Since $\mathrm{P}1$ is convex in $\{t_{k,m}\}$ with $\{p_{k,m}\}$ fixed (vice versa), according to the Karush-Kuhn-Tucker condition \cite{opt1}, the optimal $\{t^*_{k,m},p^*_{k,m}\}$ and $\{\zeta_{k}^*,\varphi^*,\chi_{k,m}^*,\theta_{k,m}^*,\xi_{k,m}^*\}$
must satisfy
\begin{subequations}
\begin{align}
&\theta^*_{k,m}t^*_{k,m}=0,~\xi^*_{k,m}p^*_{k,m}=0,~~\forall k,m, \label{slackness}
\\
&
(2-\mu)t_{k,m}^*-\frac{\zeta^*_{k}t^*_{k,m}}{T\mathrm{ln}2}\cdot\frac{v_m^*A_{k,m}}{1+v_m^*A_{k,m}p_{k,m}^*}-\xi^*_{k,m}
\nonumber\\
&
=0,~~\forall k,m, \label{kkt1}
\\
&
\zeta^*_{k}\mathrm{log}_2\left(1+v_m^*A_{k,m}p^*_{k,m}\right)-(2-\mu)Tp_{k,m}^*
\nonumber\\
&
=T\left[\varphi^*-\theta^*_{k,m}+\chi_{k,m}^*\left(1-v_m^*\right)\right],~~\forall k,m.
\label{kkt2}
\end{align}
\end{subequations}
Putting $t^*_{i,j},t^*_{i,n},p_{i,j}^*,p_{i,n}^*\neq 0$ into \eqref{slackness}, we have
\begin{align}
&\theta^*_{i,j}=\theta^*_{i,n}=\xi^*_{i,j}=\xi^*_{i,n}=0. \label{zero}
\end{align}
Further putting $\xi^*_{i,n}=0$ from \eqref{zero} into \eqref{kkt1}, the following equation is obtained
\begin{align}\label{pim}
&p^*_{i,n}=\frac{\zeta_i^*}{\mathrm{ln}2\cdot (2-\mu)T}-\frac{1}{v_n^*A_{i,n}}.
\end{align}
Substituting \eqref{pim}, $\theta^*_{i,n}=0$ from \eqref{zero}, and $v_{n}^*=1$ (due to \eqref{notvisit} and $t^*_{i,n}\neq 0$) into \eqref{kkt2}, equation \eqref{kkt2} is reformulated as
$F(v_n^*A_{i,n})=T\varphi^*$, where
\begin{align}
&F(x)=
\zeta^*_{i}\mathrm{log}_2\left[\frac{\zeta_i^*x}{\mathrm{ln}2\cdot (2-\mu)T}\right]
+\frac{(2-\mu)T}{x}
-\frac{\zeta_i^*}{\mathrm{ln}2}
\end{align}
with $x\neq 0$.
Notice that due to $v_n^*A_{i,n}\neq 0$ (as $v_n^*=1$), $F(v_n^*A_{i,n})$ is well-defined.
Similarly, by using $\xi^*_{i,j}=\theta^*_{i,j}=0$ from \eqref{zero}, we obtain $F(v_j^*A_{i,j})=T\varphi^*$. Therefore,
\begin{align}
&F(v_n^*A_{i,n})=F(v_j^*A_{i,j}). \label{n=1}
\end{align}
Now, the derivative of $F(x)$ can be computed to be
\begin{align}
\nabla_xF=&
\frac{1}{x}\left[\frac{\zeta^*_{i}}{\mathrm{ln}2}-\frac{(2-\mu)T}{x}\right]
\nonumber\\
=&\frac{1}{x}\left[(2-\mu)Tp^*_{i,n}+(2-\mu)T\left(\frac{1}{v_n^*A_{i,n}}-\frac{1}{x}\right)\right],
\end{align}
where the second equality is obtained from \eqref{pim}.
Since $p^*_{i,n}>0$, it is clear that $\nabla_xF(x)>0$ for any $x\in[v^*_nA_{i,n},v^*_jA_{i,j}]$ and $F(x)$ is a strictly increasing function of $x$ over this interval.
Combining the result from \eqref{n=1}, we have $v^*_nA_{i,n}=v_j^*A_{i,j}=\mathop{\mathrm{max}}_{l\in\mathcal{V}}~v^*_lA_{i,l}$.
This contradicts to $n\neq \mathop{\mathrm{argmax}}_{l\in\mathcal{V}}~v_l^*A_{i,l}$.
Therefore, $t^*_{i,n}=0$ at vertex $n\neq \mathop{\mathrm{argmax}}_{l\in\mathcal{V}}~v_l^*A_{i,l}$.

\section{Transformation from $\mathrm{P}2$ to $\mathrm{P}3$ with Fixed $\mathbf{v}=\widetilde{\mathbf{v}}$}

With $q_k$ in \eqref{qk}, the first constraint of $\mathrm{P}2$ is always satisfied.
Therefore, we can re-write $\mathrm{P}2$ as
\begin{align}
\mathop{\mathrm{min}}_{\substack{\{s_{k}>0,\mathbf{W},\lambda_m\}}}
~~&\mu\left(\alpha_1/a+\alpha_2\right)\mathrm{Tr}(\mathbf{D}^{T}\mathbf{W})
\nonumber\\
&
+(2-\mu)\sum_{k=1}^K\frac{s_k\left(2^{\gamma_k/s_k}-1\right)}{B_k(\widetilde{\mathbf{v}})}
\nonumber\\
~~~~\mathrm{s.t.}~~~~~~~~&\sum_{k=1}^Ks_{k}\leq T-
\frac{1}{a}\mathrm{Tr}(\mathbf{D}^{T}\mathbf{W}),
\nonumber\\
&
\eqref{P1c}-\eqref{edge}.  \label{Appendix B P}
\end{align}
Since $s_k$ is not involved in $\eqref{P1c}-\eqref{edge}$, the partial Lagrangian of the problem \eqref{Appendix B P} with respect to $\{s_k\}$ is given by
\begin{align}
&\mathcal{L}\left(\{s_{k}\},\varphi,\{\theta_{k}\} \right)=
(2-\mu)\sum_{k=1}^K\frac{s_k\left(2^{\gamma_k/s_k}-1\right)}{B_k(\widetilde{\mathbf{v}})}
\nonumber\\
&+\varphi\left[\sum_{k=1}^Ks_{k}-T+\frac{1}{a}\mathrm{Tr}(\mathbf{D}^{T}\mathbf{W})\right]
-\sum_{k=1}^K\theta_{k}s_{k},
 \nonumber
\end{align}
where $\varphi\geq 0$ and $\theta_{k}\geq 0$ are Lagrange multipliers.
Since problem \eqref{Appendix B P} is convex in $\{s_k\}$, we must have $\frac{\partial \mathcal{L}}{\partial s_k}=0$ according to the KKT condition,
and the optimal $\{\widehat{s}_{k}\}$ and $(\widehat{\varphi},\widehat{\theta}_k)$ should together satisfy
\begin{align}\label{B1}
&\frac{(2-\mu)}{B_k(\widetilde{\mathbf{v}})}\cdot
\left(1+\mathrm{ln}2\cdot 2^{\gamma_k/\widehat{s}_{k}}\cdot\gamma_k/\widehat{s}_{k}-2^{\gamma_k/\widehat{s}_{k}}\right)
=\widehat{\varphi}-\widehat{\theta}_k.
\end{align}
Since the derivative of the left hand side of \eqref{B1} with respect to $\widehat{s}_{k}$ is
$-\frac{(2-\mu)}{B_k(\widetilde{\mathbf{v}})}\cdot
\mathrm{ln}^22\cdot 2^{\gamma_k/\widehat{s}_{k}}\cdot\gamma_k^2/(\widehat{s}_{k})^3<0,
$
the left hand side function in \eqref{B1} is a strictly decreasing function of $\widehat{s}_{k}$.
Furthermore, due to
\begin{align}
&\mathop{\mathrm{lim}}_{\widehat{s}_{k}\rightarrow+\infty}\frac{(2-\mu)}{B_k(\widetilde{\mathbf{v}})}\cdot
\left(1+\mathrm{ln}2\cdot 2^{\gamma_k/\widehat{s}_{k}}\cdot\gamma_k/\widehat{s}_{k}-2^{\gamma_k/\widehat{s}_{k}}\right)
=0,
\end{align}
it is clear that $\widehat{\varphi}-\widehat{\theta}_k>0$.
As $\widehat{\theta}_{k} \widehat{s}_k=0$ from the complementary slackness condition, and due to $\widehat{s}_k>0$,
we must have $\widehat{\theta}_{k}=0$, and therefore $\widehat{\varphi}>0$.

Finally, from $\widehat{\varphi}>0$ and the complementary slackness condition
$\widehat{\varphi}\left[\frac{1}{a}\mathrm{Tr}(\mathbf{D}^{T}\widehat{\mathbf{W}})+\sum_{k=1}^K\widehat{s}_{k}-T\right]=0$,
the equation $\mathrm{Tr}(\mathbf{D}^{T}\widehat{\mathbf{W}})=a\left(T-\sum_{k=1}^K\widehat{s}_{k}\right)$ holds.
Substituting this result into $\mathrm{P}2$, $\mathrm{P}2$ becomes
\begin{align}\label{B2}
\mathop{\mathrm{min}}_{\substack{\{s_{k}>0,\mathbf{W},\lambda_m\}}}~
&\mu\left(\alpha_1+\alpha_2a\right)\left(T-\sum_{k=1}^Ks_{k}\right)
\nonumber\\
&
+(2-\mu)\sum_{k=1}^K\frac{s_k}{B_k(\widetilde{\mathbf{v}})}\left(2^{\gamma_k/s_k}-1\right),
\nonumber\\
\mathrm{s.t.}~~~~~~&\sum_{k=1}^Ks_{k}=T-
\frac{1}{a}\mathrm{Tr}(\mathbf{D}^{T}\mathbf{W}),
\nonumber\\
&
\eqref{P1c}-\eqref{edge}.
\end{align}
Now the objective function of \eqref{B2} is independent of $\mathbf{W}$ and is a decreasing function of $s_k$ (since its derivative can be shown to be negative).
Therefore, the minimum value of the objective function in \eqref{B2} is obtained when $s_k$ is maximized.
Since $\sum_{k=1}^Ks_{k}=T-\mathrm{Tr}(\mathbf{D}^{T}\mathbf{W})/a$, maximizing $T-\mathrm{Tr}(\mathbf{D}^{T}\mathbf{W})/a$ helps in enlarging the maximum values of $\{s_k\}$. Therefore, \eqref{B2} is equivalently transformed into $\mathrm{P}3$.

\section{Proof of Proposition 2}

We first prove the left part of equation \eqref{bound}.
In particular, since $\Lambda(x)$ is a convex function (see Fig. 3), we must have
\begin{align}
&
\Lambda\left(\frac{1}{\sum_{k=1}^K\gamma_k}\sum_{k=1}^K\gamma_k\cdot\frac{B_k(\widetilde{\mathbf{v}})\widehat{\rho}}{2-\mu}\right)
\nonumber\\
&
\leq
\frac{1}{\sum_{k=1}^K\gamma_k}
\sum_{k=1}^K\gamma_k\cdot
\Lambda\left(\frac{B_k(\widetilde{\mathbf{v}})\widehat{\rho}}{2-\mu}\right)
=\frac{\Upsilon(\widetilde{\mathbf{v}})}{\sum_{k=1}^K\gamma_k}, \label{Appendix C 1}
\end{align}
where ``$\leq$'' is due to Jensen's inequlity and ``$=$'' is due to \eqref{rho}.
By further applying function $-\nabla\Theta$ to both sides of \eqref{Appendix C 1}, and since $-\nabla\Theta$ is a strictly decreasing function,
 \eqref{Appendix C 1} becomes
\begin{align}
&\frac{1}{\sum_{k=1}^K\gamma_k}\sum_{k=1}^K\gamma_kB_k(\widetilde{\mathbf{v}})\cdot
\frac{\widehat{\rho}}{2-\mu}
\geq-\nabla\Theta\left(\frac{\Upsilon(\widetilde{\mathbf{v}})}{\sum_{k=1}^K\gamma_k}\right), \nonumber
\end{align}
which immediately leads to the left part of \eqref{bound}.

Next, we prove the right part of equation \eqref{bound}.
More specifically, since $B_k(\widetilde{\mathbf{v}})\geq\mathrm{min}_l~B_l(\widetilde{\mathbf{v}})$ and $\widehat{\rho}\geq0$, we have
\begin{align}
&\frac{B_k(\widetilde{\mathbf{v}})\widehat{\rho}}{2-\mu}
\geq\frac{\mathrm{min}_l~B_l(\widetilde{\mathbf{v}})\widehat{\rho}}{2-\mu}.  \label{Appendix C 2}
\end{align}
Applying $\Lambda(x)$ to both sides of \eqref{Appendix C 2}, and since $\Lambda(x)$ is a strictly decreasing function,
 \eqref{Appendix C 2} becomes
\begin{align}
&\Lambda\left(\frac{B_k(\widetilde{\mathbf{v}})\widehat{\rho}}{2-\mu}\right)
\leq
\Lambda\left(\frac{\mathrm{min}_l~B_l(\widetilde{\mathbf{v}})\widehat{\rho}}{2-\mu}\right).
\end{align}
Based on the above result and equation \eqref{rho}, it is clear that
\begin{align}
&\sum_{k=1}^K\gamma_k\cdot
\Lambda\left(\frac{\mathrm{min}_l~B_l(\widetilde{\mathbf{v}})\widehat{\rho}}{2-\mu}\right)
\geq\Upsilon(\widetilde{\mathbf{v}}),
\end{align}
which is equivalent to
\begin{align}
&
\frac{\mathrm{min}_l~B_l(\widetilde{\mathbf{v}})\widehat{\rho}}{2-\mu}
\leq
-\nabla\Theta\left(\frac{\Upsilon(\widetilde{\mathbf{v}})}{\sum_{k=1}^K\gamma_k}\right).
\end{align}
This leads to the right part of \eqref{bound}, and the proof is completed.

\section{Proof of Lower Bound Property of $\Psi$ in \eqref{Psi}}

To begin with, it is noticed that $\Xi\left(\widetilde{\mathbf{v}}\right)$ in \eqref{Xi} is the optimal value of problem $\mathrm{P}3$.
Therefore, to prove $\Psi(\mathcal{F}_{(z_1,\cdots,z_N)})\leq \Xi(\widetilde{\mathbf{v}})$ for any $\widetilde{\mathbf{v}}\in\mathcal{F}_{(z_1,\cdots,z_N)}$, we consider two relaxations for $\mathrm{P}3$ with $\widetilde{\mathbf{v}}\in\mathcal{F}_{(z_1,\cdots,z_N)}$.

(i) Based on the definition $B_k(\widetilde{\mathbf{v}}):=\mathop{\mathrm{max}}_{l}~\widetilde{v}_l A_{k,l}$, it can be seen that
$B_k(\mathbf{v}_1)\geq B_k(\mathbf{v}_2)$ if $\mathbf{v}_1\succeq \mathbf{v}_2$, where $\succeq$ means ``Pareto dominance''.
Furthermore, since $[\mathbf{z}^{T},\mathbf{1}_{M-N}^{T}]^{T}\succeq \widetilde{\mathbf{v}}$ for any $\widetilde{\mathbf{v}}\in\mathcal{F}_{(z_1,\cdots,z_N)}$, we have
$B_k\left([\mathbf{z}^{T},\mathbf{1}_{M-N}^{T}]^{T}\right)\geq B_k(\widetilde{\mathbf{v}})$ for any $\widetilde{\mathbf{v}}\in\mathcal{F}_{(z_1,\cdots,z_N)}$.
Using this result, the objective function of $\mathrm{P}3$ can be lower bounded by $\mu\left(\alpha_1+\alpha_2a\right)\left(T-\sum_{k=1}^Ks_{k}\right)
+(2-\mu)\sum_{k=1}^K\frac{\gamma_k}{B_k\left([\mathbf{z}^{T},\mathbf{1}_{M-N}^{T}]^{T}\right)}
\Theta\left(\frac{s_k}{\gamma_k}\right)$.

(ii) Notice that
\begin{align}
&\mathop{\mathrm{max}}_{\substack{\mathbf{W},\{\lambda_m\}}} \Big\{T-\frac{1}{a}\mathrm{Tr}(\mathbf{D}^{T}\mathbf{W}):\eqref{P1c}-\eqref{edge},\widetilde{\mathbf{v}}\in\mathcal{F}_{(z_1,\cdots,z_N)}\Big\}
\nonumber\\
&
\leq \mathop{\mathrm{max}}_{\substack{\mathbf{W}}} \Big\{T-\frac{1}{a}\mathrm{Tr}(\mathbf{D}^{T}\mathbf{W}):\eqref{P1c},\eqref{edge},\widetilde{\mathbf{v}}\in\mathcal{F}_{(z_1,\cdots,z_N)}\Big\}
\nonumber\\
&
=\Phi(\mathbf{z}), \label{D1}
\end{align}
where the inequality is obtained by dropping constraints \eqref{subtour1} and \eqref{subtour2}.
Using \eqref{D1}, the constraint of $\mathrm{P}3$ can be relaxed into
$\sum_{k=1}^Ks_k\leq \Phi(\mathbf{z})$.

Based on the above two relaxations, the problem $\mathrm{P}3$ with $\widetilde{\mathbf{v}}\in\mathcal{F}_{(z_1,\cdots,z_N)}$ is relaxed into
\begin{align}
\mathop{\mathrm{min}}_{\substack{\{s_{k}>0\}}}
~~&\mu\left(\alpha_1+\alpha_2a\right)\left(T-\sum_{k=1}^Ks_{k}\right)
\nonumber\\
&
+(2-\mu)\sum_{k=1}^K\frac{\gamma_k}{B_k\left([\mathbf{z}^{T},\mathbf{1}_{M-N}^{T}]^{T}\right)}
\Theta\left(\frac{s_k}{\gamma_k}\right)
\nonumber\\
~~\mathrm{s.t.}~~~~&\sum_{k=1}^Ks_k\leq \Phi(\mathbf{z}). \label{58}
\end{align}
Using the result from \eqref{sk}-\eqref{rho}, the optimal $\{s_k\}$ to the problem \eqref{58} is given by
\begin{align}\label{59}
&s_k^*=\gamma_k\cdot\Lambda\left(\frac{B_k\left([\mathbf{z}^{T},\mathbf{1}_{M-N}^{T}]^{T}\right)\delta}{2-\mu}\right),
\end{align}
with $\delta$ obtained from \eqref{kappa}.
Putting \eqref{59} into the objective function of \eqref{58}, we immediately obtain $\Psi$ in \eqref{Psi}, which is obviously a lower bound to the objective function $\Xi\left(\widetilde{\mathbf{v}}\right)$ of $\mathrm{P}3$ for any $\widetilde{\mathbf{v}}\in\mathcal{F}_{(z_1,\cdots,z_N)}$.

\end{document}